\documentclass[preprint,11pt,3p,oneside]{elsarticle} 
\journal{arXiv} 

\usepackage{lineno}

\usepackage[hidelinks]{hyperref}
\usepackage{verbatim} 
\usepackage{graphicx}
\usepackage{svg}
\usepackage{amsmath}   
\usepackage[utf8]{inputenc}
\usepackage{enumitem}
\usepackage{float}    
\usepackage{subcaption} 
\usepackage{caption}  
\usepackage{tabularx}
\usepackage{booktabs}
\usepackage{makecell}
\usepackage{tikz}
\usepackage{siunitx}
\usetikzlibrary{shapes,arrows,positioning}
\biboptions{sort&compress}

\begin{document}

\begin{frontmatter}
    
\title{High-accuracy ultrasonic positioning of calibration sources in the Jiangmen Underground Neutrino Observatory}
\author[1]{Ziqian Xiang\fnref{fn1}}
\ead{xiangziqian@sjtu.edu.cn}
\author[2]{Rongcheng Chen}
\author[1]{Zhangmin Chen}
\author[2]{Qian Chen}
\author[1]{Diwash Ghimire}
\author[3]{Jiaqi Hui}
\author[1]{Junting Huang}
\author[1]{Junjie Jiang}
\author[2]{Daijin Li}
\author[1]{Haojing Lai}
\author[2]{Kai Luo}
\author[1]{Rui Li}
\author[1]{Yilin Liao}
\author[3,4,5]{Jianglai Liu}
\author[1,6]{Yue Meng\corref{cor1}}
\ead{mengyue@sjtu.edu.cn}
\author[2]{Yazhen Shi}
\author[2]{Duo Teng}
\author[2]{Linwei Tao}
\author[2]{Qi Wang}
\author[2]{Changsheng Ye}
\author[2]{Guolei Zhu}
\author[1]{Ping Zhang}
\author[1]{Tao Zhang}

\fntext[fn1]{First author.}
\cortext[cor1]{Corresponding author.}

\affiliation[1]{
                department={School of Physics and Astronomy,},
                organization={Shanghai Jiao Tong University},
                postcode={200240},
                city={Shanghai},
                country={China}}

\affiliation[2]{
                department={School of Marine Science and Technology,},
                organization={Northwestern Polytechnical University},
                postcode={710072},
                city={Xi'an},
                country={China}}

\affiliation[3]{
                department={Tsung-Dao Lee Institute,},
                organization={Shanghai Jiao Tong University},
                postcode={201210},
                city={Shanghai},
                country={China}}
\affiliation[4]{
                department={School of Physics and Astronomy, MOE Key Laboratory for Particle Astrophysics and Cosmology, Shanghai Key Laboratory for Particle Physics and Cosmology,},
                organization={Shanghai Jiao Tong University},
                postcode={200240},
                city={Shanghai},
                country={China}}
\affiliation[5]{
                department={New Cornerstone Science Laboratory, Tsung-Dao Lee Institute,},
                organization={Shanghai Jiao Tong University},
                postcode={201210},
                city={Shanghai},
                country={China}}
\affiliation[6]{
                organization={Shanghai Jiao Tong University Sichuan Research Institute},
                postcode={610213},
                city={Chengdu},
                country={China}}

\begin{abstract}
Precise source positioning is essential for detector calibration in large liquid scintillator detectors such as JUNO, particularly in regions where purely mechanical control is insufficient. An ultrasonic positioning system has been developed to reconstruct the three-dimensional coordinates of a calibration source without interfering with photon collection or contaminating the liquid scintillator. The method combines a sound-speed modeling based on dedicated laboratory measurements and in-detector temperature profiles, waveform-based arrival-time reconstruction, and an in-situ calibration of the effective receiver geometry using central-axis deployments. With six active receivers, central-axis positioning yields a mean error of 1.23 cm relative to the known deployment reference. For off-axis operation in the Cable Loop System calibration plane, a detector-realistic simulation that includes timing resolution, sound-speed variation, and receiver-coordinate smearing predicts a positioning uncertainty of 2.40 cm. These results demonstrate that ultrasonic positioning can provide centimetre-level source accuracy for large liquid scintillator detectors and can support off-axis calibration in JUNO-like experiments.
\end{abstract}


\end{frontmatter}

\section{Introduction}

Jiangmen Underground Neutrino Observatory (JUNO) is a large liquid scintillator neutrino experiment designed for precision measurements of reactor antineutrino oscillations and for determining the neutrino mass ordering~\cite{An_2016, abusleme2025measurementreactorneutrinooscillations}. To achieve its physics goals, the detector response must be calibrated accurately throughout the full target volume. In particular, spatial non-uniformity in light collection requires radioactive calibration sources to be deployed at well-defined positions. Compared with earlier neutrino experiments, in which source deployment was often limited to the detector central axis, calibration in JUNO, by design, extends to off-axis regions in order to map the detector response over a much larger volume. The central detector of JUNO is a 17.7~m-radius acrylic sphere filled with liquid scintillator (LS), and the signal is detected by the Photomultiplier tubes (PMTs). Therefore, the calibration source must be deployed using hardware that satisfies background and compatibility requirements and introduces minimal optical shadowing. Meanwhile, the only access to the acrylic ball is through the central chimney, which limits the choices of deployment.

\begin{figure}[hbt!]
    \centering
    \includegraphics[width=0.5\linewidth]{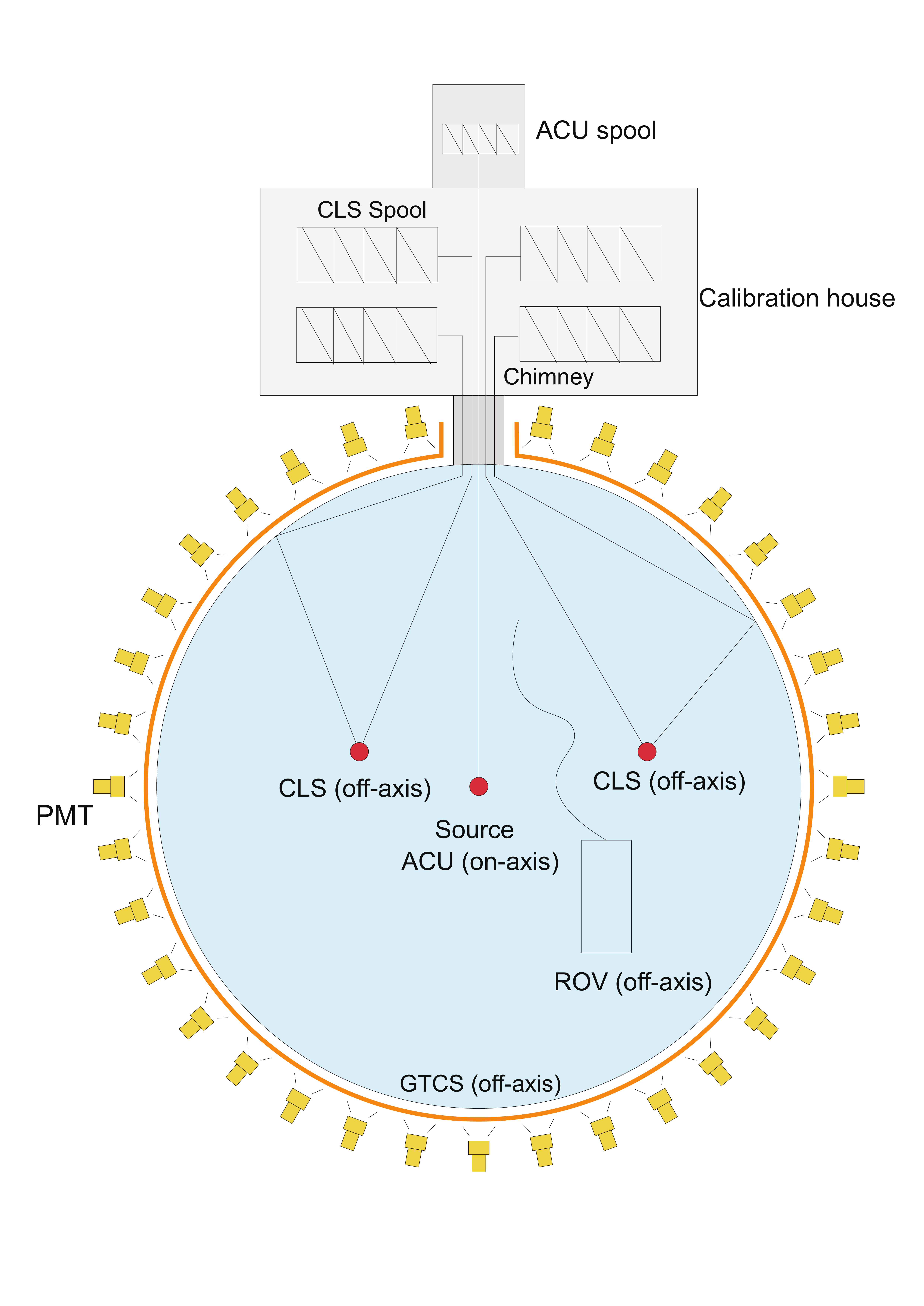}
    \caption{Schematic diagram of the JUNO central detector calibration systems. 
    The PMT array is arranged around the acrylic sphere filled with liquid scintillator. Three calibration subsystems are illustrated: the ACU, which is deployed along the central on-axis vertical line; the CLS operated in off-axis positions; the ROV, providing off-axis calibration capabilities; and the GTCS, enabling calibration in the boundary of the acrylic. The Calibration House located at the top of the detector houses the winch spools (ACU Spool and CLS Spool) and the Chimney for deployment. The lines indicate the cable routing paths for the respective sources.}
    \label{fig:calibration_system_schematic}
 
\end{figure}

With these considerations, JUNO established a comprehensive in-situ calibration program consisting of four complementary systems (Fig.~\ref{fig:calibration_system_schematic}): the Automatic Calibration Unit (ACU), the Cable Loop System (CLS), the Remotely Operated Vehicle (ROV), and a peripheral guide-tube calibration system (GTCS) mounted outside the acrylic sphere~\cite{Feng_2018, Zhang:2020grf, Hui_2025, Hui:2021dnh, JUNO:2020xtj}. These systems provide one-, two-, and three-dimensional source deployment capabilities in different detector regions. Their positioning principles and constraints differ significantly. The central-axis ACU deployment provides intrinsically precise positioning through cable metrology, and the guide-tube system relies on dedicated position sensors since it is outside the CD. By contrast, the two off-axis deployments inside the acrylic sphere, the CLS and the ROV, require an independent positioning method. Considering the strict constraints on acrylic ball access, contamination, optical shadowing and accuracy requirement, JUNO adopts acoustic-based~\cite{Zhang:2020grf} and CCD-based~\cite{Li2022JUNO} positioning scheme for these configurations.

In this work, we focus on the ultrasonic positioning system (USS), which determines the three-dimensional coordinates of a calibration source from the time of flight of acoustic signals. The system consists of one ultrasonic emitter attached to the source and multiple receivers mounted on the inner surface of the acrylic sphere~\cite{sym16091218, Zhu:2019vay}. In principle, the source position is reconstructed through a global fit that evaluates the measured time-of-flight values against the expected acoustic path lengths based on known receiver coordinates and the speed of sound. The implementation is not straightforward, as it faces two coupled experimental challenges in JUNO. First, the speed of sound in LS cannot be taken directly from a commercial water-calibrated instrument without introducing uncontrolled systematic uncertainty, and must instead be calibrated specifically for the LS. Second, the receiver coordinates used in the reconstruction are expected to be expressed in the PMT reference frame, while small deformations after LS filling in the acrylic may lead to residual misalignments between the surveyed USS reference frame and the PMT reference frame. These considerations make both an LS-specific sound-speed measurement and an in-situ calibration of the full positioning chain essential.

This work addresses the above two coupled challenges in ultrasonic source positioning for JUNO. To this end, we combine dedicated laboratory measurements with LS temperature profiles in the JUNO acrylic ball, calibrate the receiver coordinates with central-axis deployments, and evaluate the resulting positioning performance for the CLS configuration under simulated USS conditions.
\begin{figure}[!hbt]
    \centering
    \includegraphics[width=0.7\linewidth]{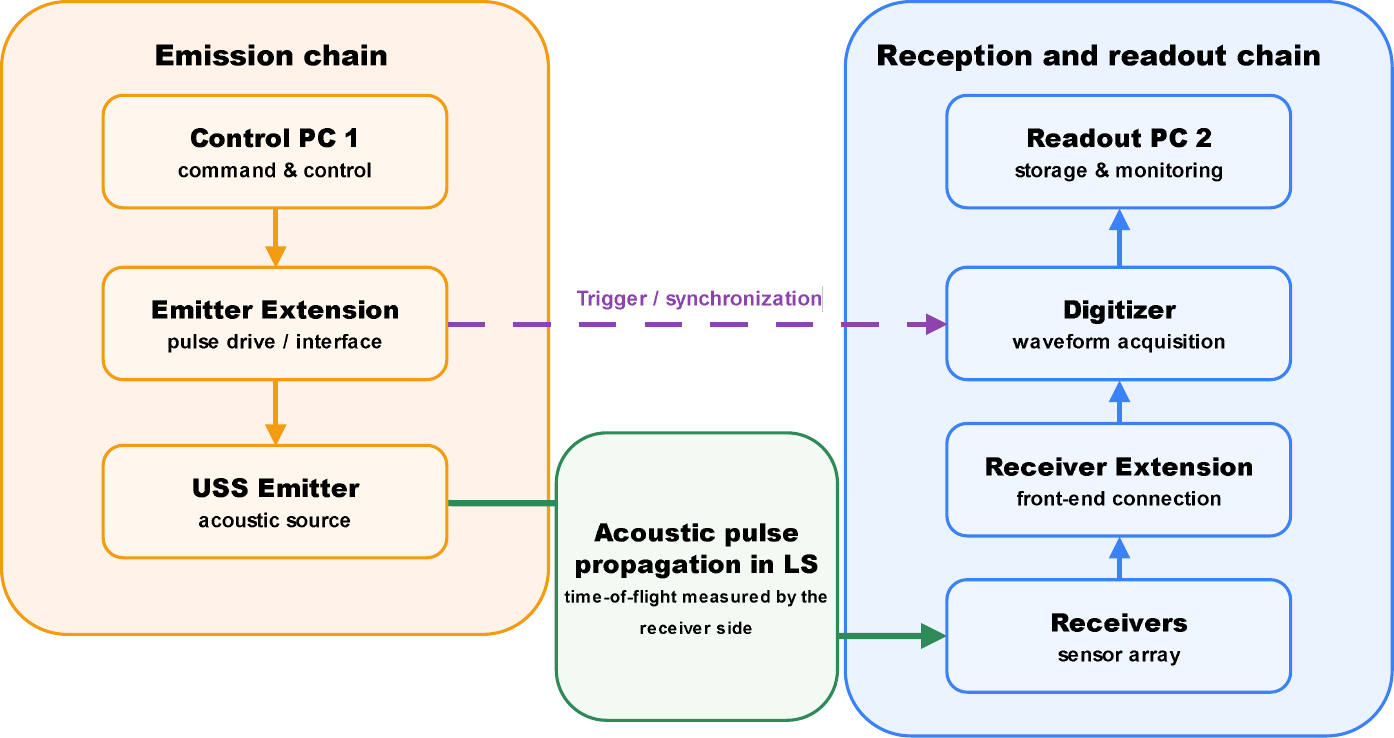}
    \caption{Schematic diagram of the hardware workflow of the USS. Amplified electrical signal driven by the emitter extension excites the ultrasonic emitter to generate acoustic waves in the LS. The propagated signals are detected by the receivers, processed by the readout electronics, and digitized by the digitizer, of which the time window is aligned to the trigger signal provided by the emitter extension with a measurable time offset.}
    \label{fig:hardware}
\end{figure}
\section{System Design}

The overall workflow of the system, including signal emission, reception, and readout, is shown in Fig.~\ref{fig:hardware}. 
The hardware modules were developed by Northwestern Polytechnical University, including a host computer, an emitter extension, a receiver extension, ultrasonic emitters, and ultrasonic receivers~\cite{Zhu:2019vay}. The digitizer is developed by Shanghai Jiao Tong University. It records 62000 sampling points of 0.512~µs after detecting a TTL signal in the trigger channel.

\begin{figure}[!hbt]

\centering
    \centering
    \begin{subfigure}{0.2\textwidth}
        \centering
        \includegraphics[width=1\linewidth]{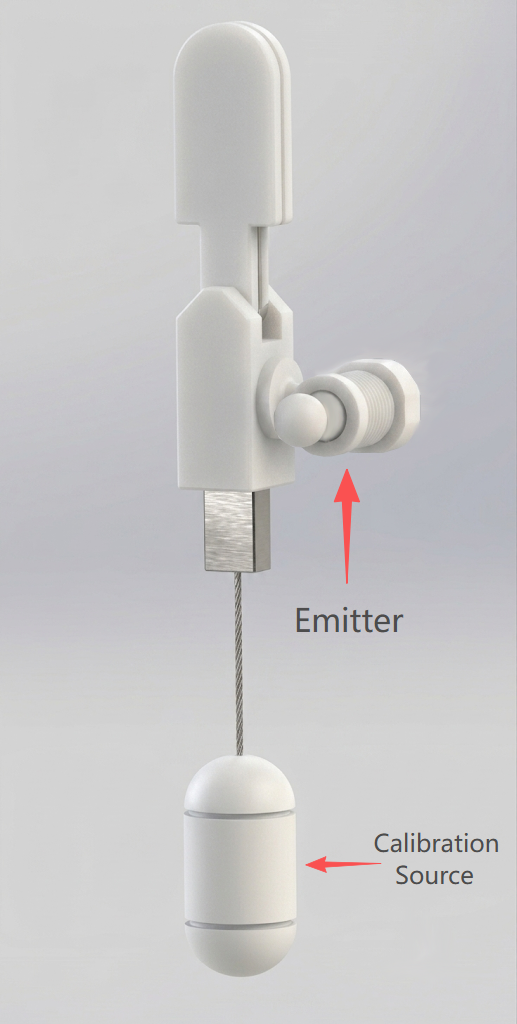}
        \caption{}
        \label{fig:emitter}
    \end{subfigure}
    \hfill
    \begin{subfigure}{0.2\textwidth}
        \centering
        \includegraphics[width=1\linewidth]{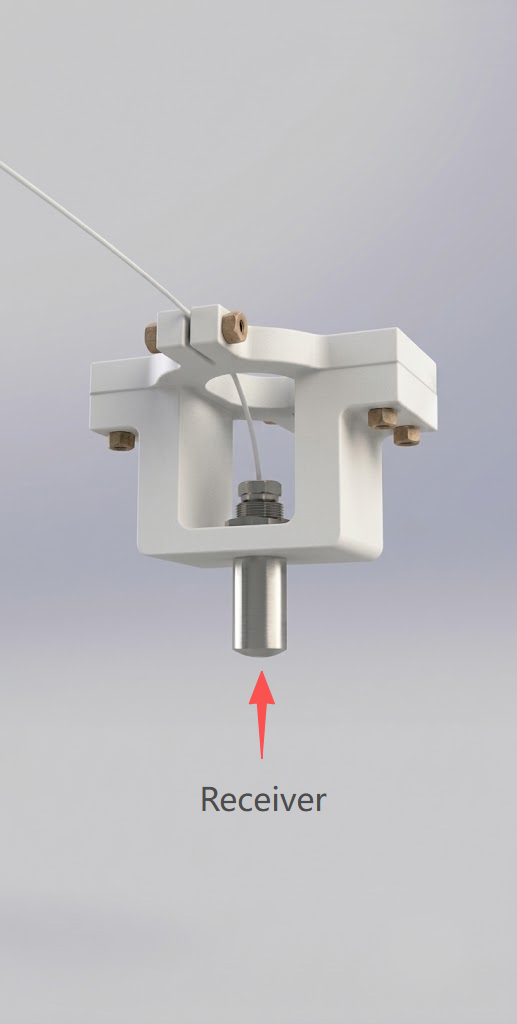}
        \caption{}
        \label{fig:receiver}
    \end{subfigure}
    \hfill
    \begin{subfigure}{0.454\textwidth}
        \centering
        \includegraphics[width=1\linewidth]{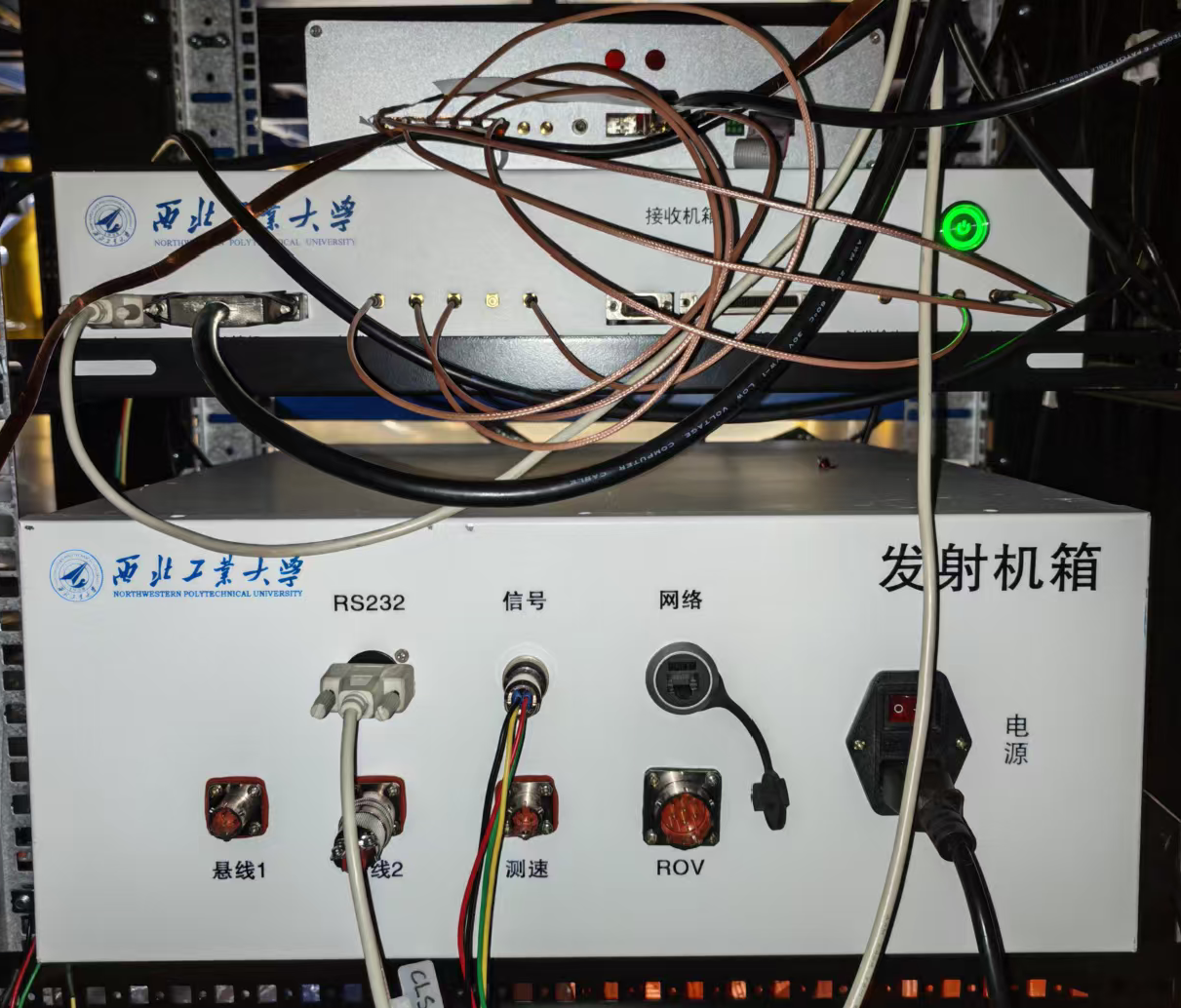}
        \caption{}
        \label{fig:controlbox}
    \end{subfigure}
    \caption{Hardware components of the USS: 
(a) Custom mounting fixture and ultrasonic receiver assembly designed for attachment to the inner surface of the acrylic sphere; 
(b) Ultrasonic emitter assembly coupled with the calibration source and the CLS steering cables; 
(c) Integrated instrumentation and readout rack, comprising the compact digitizer (top), the multi-channel receiver readout chassis (middle), and the emitter control box (bottom).}\label{fig::hardwareSets}
 
\end{figure}

\begin{figure}[!hbt]
\centering
\includegraphics[width=0.45\textwidth]{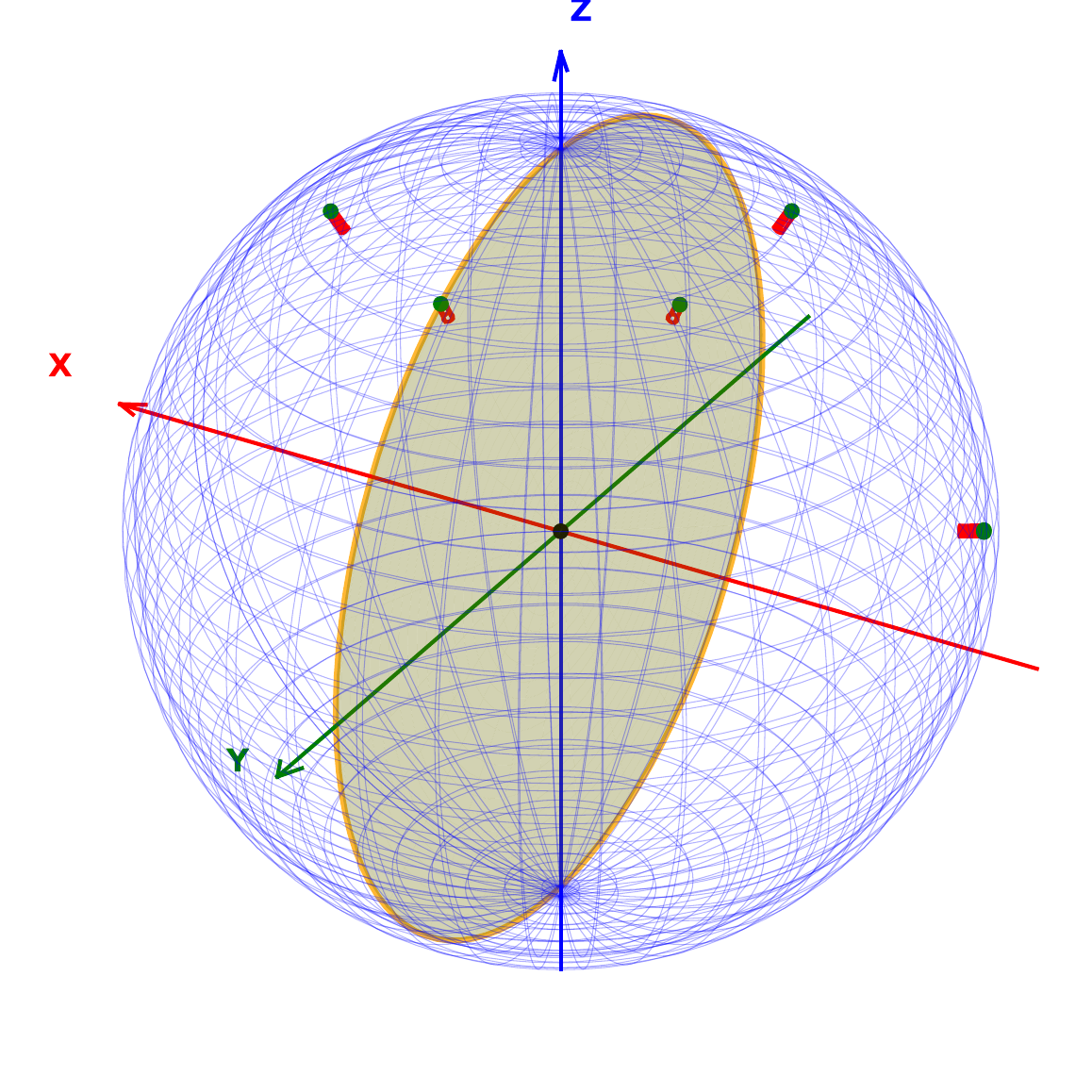}

    \centering
    \begin{subfigure}{0.45\textwidth}
        \centering
        \includegraphics[width=1\linewidth]{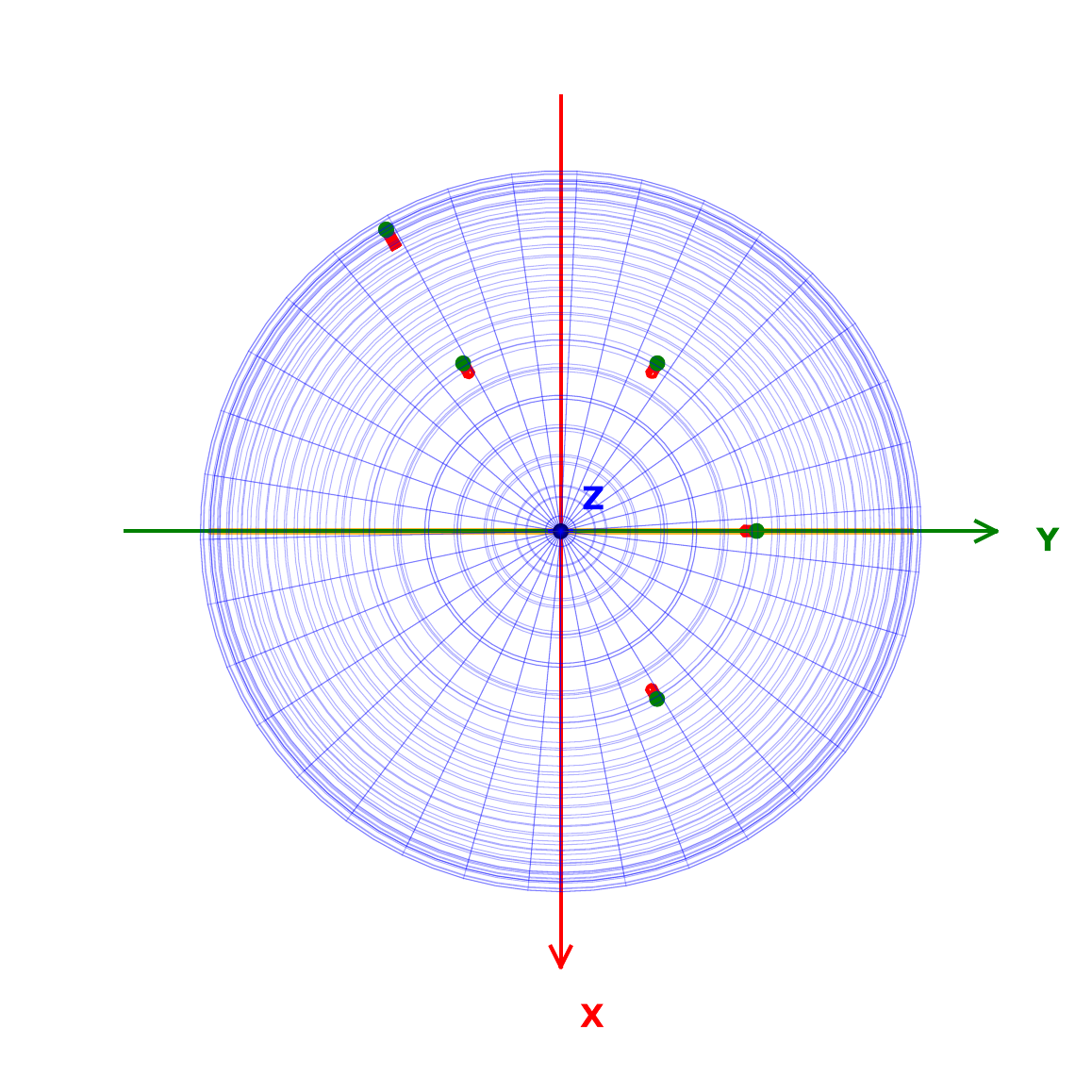}
        \caption{}
        \label{fig:subfig1}
    \end{subfigure}
    \hfill
    \begin{subfigure}{0.45\textwidth}
        \centering
        \includegraphics[width=1\linewidth]{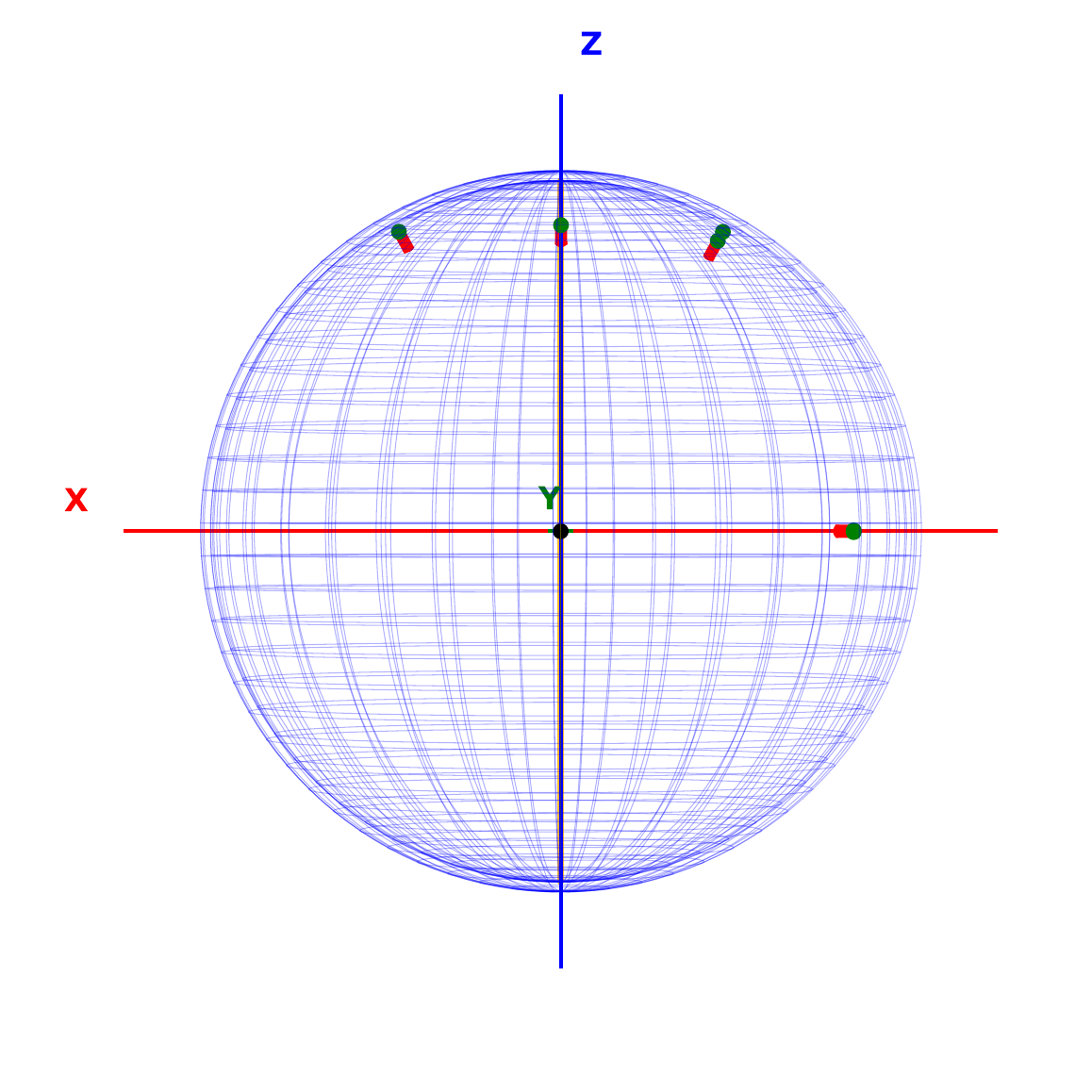}
        \caption{}
        \label{fig:subfig2}
    \end{subfigure}
    
\caption{Geometry of the JUNO central detector, the USS receiver array, and the CLS calibration plane. The blue mesh denotes the acrylic boundary, the red cylinders mark the six active USS (two overlapping at the equator in this view) receivers mounted on the inner acrylic surface, and the yellow markers indicate the CLS calibration plane; the lower panels show two projected views of the same geometry to illustrate the coverage and the relative position of the target plane and the receivers. Although ten receivers were installed originally, only six remained active after acrylic bonding, water filling, and LS filling.}\label{fig1}

\end{figure}

Different cable types are adopted for the emitter-to-controller and receivers-to-readout according to their mechanical and optical constraints. In the CLS configuration, the emitter is connected by two stainless-steel cables, which simultaneously provide mechanical support and electrical transmission for driving the emitter, as shown in Fig.~\ref{fig:emitter}. During the ACU-based calibration of the USS, the emitter is instead deployed with a single mildly load-bearing coaxial cable, since only one cable is used and the return path must therefore be integrated into the same line. By contrast, the receivers are fixed to the inner surface of the acrylic sphere by a Teflon base and do not require load-bearing cables, as shown in Fig.~\ref{fig:receiver}. To minimize optical shadowing and radioactivity, their signals are transmitted through thin coaxial cables with Teflon outer insulation~\cite{LowBG}. These cables are routed from the top chimney of the central detector to the receiver readout chassis, where the received acoustic signals are read out, shown in Fig.~\ref{fig:controlbox}. The operating platform is located between the top chimney and the calibration house. The cable lengths are approximately 100 m, and the resulting electrical transmission time is about 50 ns, which are negligible for the present analysis. The layout of the USS receiver array and the CLS calibration plane is shown in Fig.~\ref{fig1}.

The support structure in the calibration operating area is connected to the isolated stable circuit. The USS readout electronics share the same isolated stable circuit as the CD PMT system, while auxiliary equipment (e.g., ventilation and mechanical systems) is connected to a separate ground to minimize electrical noise coupling. Dedicated tests show that the operation of the USS does not produce observable PMT gain drift or dark count rate variation. The calibration system and the ultrasonic positioning system can therefore be operated simultaneously.

\section{Methodology}
The USS determines the source coordinates by minimizing the differences between the measured and geometrically expected acoustic time-of-flight. The underlying methodology relies on three essential components discussed sequentially in this section: The speed of sound in the medium, the signal propagation time, and the reconstruction of the source position.

\subsection{Sound-Speed Model}
\begin{figure}[!hbt]
    \centering
    \includegraphics[width=0.6\linewidth]{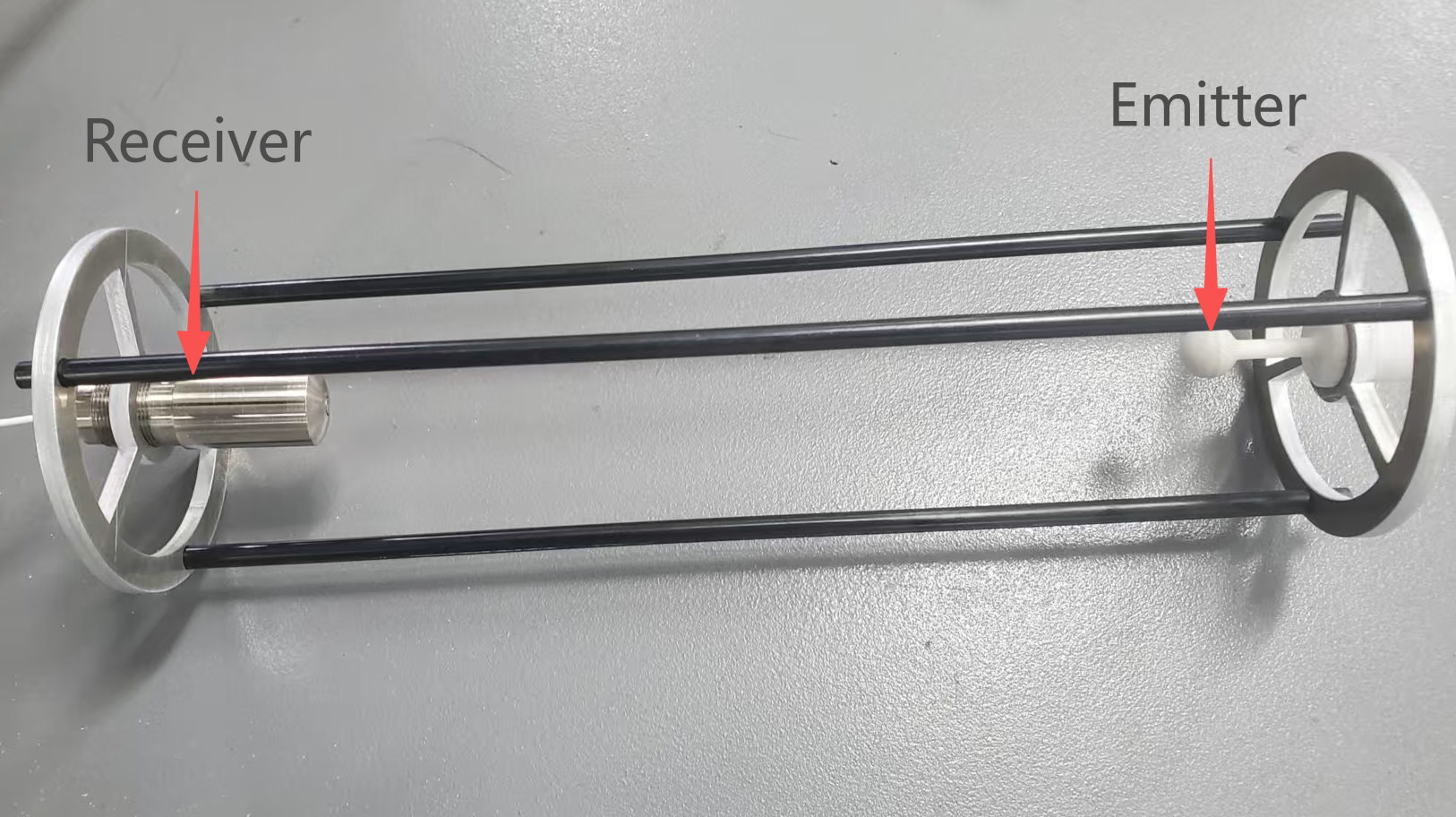}
    \caption{Photograph of the compact device developed for measuring the speed of sound in LS. The device consists of one emitter and one receiver identical to those used in JUNO. A hollow stainless-steel base and a carbon-fiber/PEEK support structure were adopted to reduce acoustic reflections and structure-borne sound transmission.}
    \label{fig:birdcage}
\end{figure}

Sound waves propagate through collective vibrations in the medium, and the corresponding phase velocity depends on the medium density and bulk modulus. In LS, these quantities may vary with pressure and temperature. In principle, the speed of sound could be measured directly in situ. In practice, however, such an approach is not suitable for the JUNO central detector. Deploying an additional measurement device inside the acrylic would be difficult because of the limited access and the requirement of precise geometrical control in a large LS volume. We therefore adopt an indirect approach based on temperature measurements in the acrylic together with dedicated laboratory measurements of the sound speed in LS. In the present analysis, we assume that the dominant variation in the sound speed arises from temperature, while the effects of pressure and composition are subdominant over the whole operating range \cite{DelGrosso1974}. For this laboratory measurements, commercial sound-speed meters are unsuitable, as they are typically calibrated for water and would introduce uncontrolled systematic uncertainties in the LS. Instead, we employed a custom-designed device (shown in Fig.~\ref{fig:birdcage}) to map the relationship between the LS temperature and sound speed, deriving the acoustic velocity directly from the measured time-of-flight across a precisely known distance.

The device was first calibrated in water with a commercial sound-speed meter in order to determine the effective distance between the emitter and receiver. Since the commercial instrument provides a high precision of $0.01\,\mathrm{m/s}$, the primary systematic uncertainty arises from thermal variations. To ensure a highly stable measurement environment, the water volume was actively regulated using a heater and a mechanical stirrer, monitored by an array of four spatially distributed temperature sensors. Under this configuration, a global temperature stability of $\pm 0.1\,^\circ\mathrm{C}$ was maintained throughout the test volume. 

\begin{figure}[!hbt]
\centering
\includegraphics[width=0.6\linewidth]{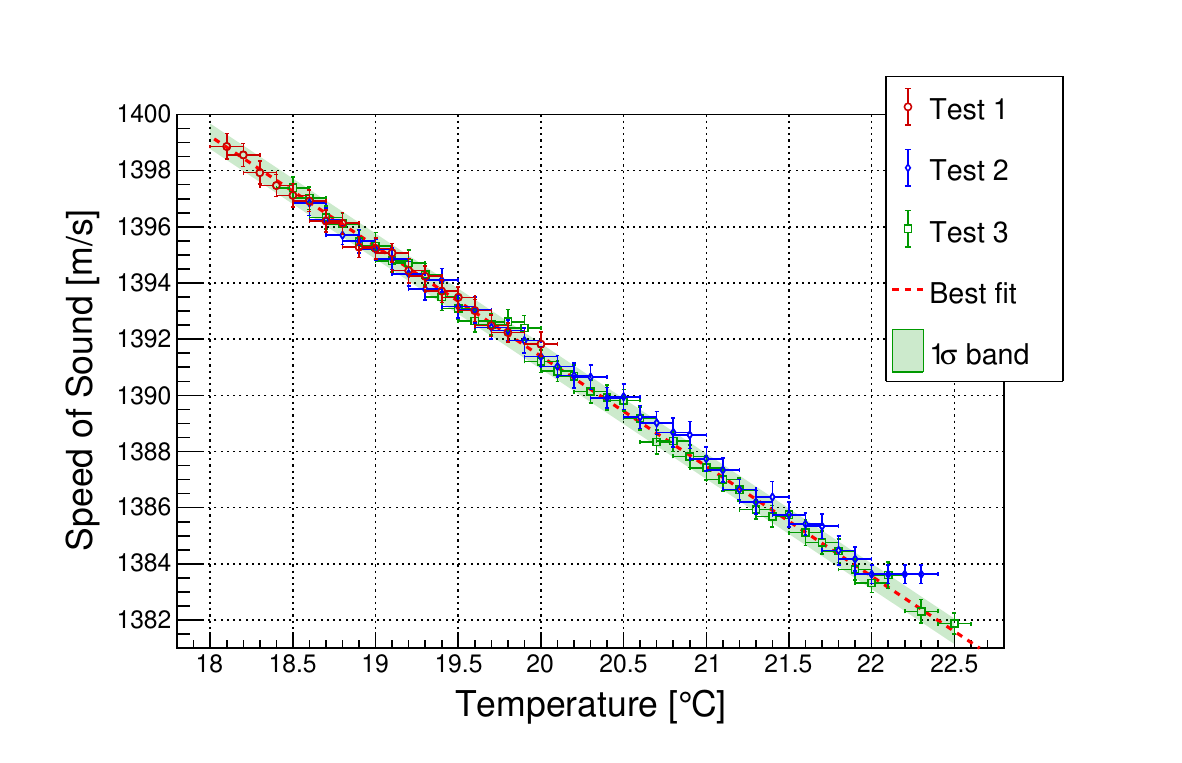}

\caption{Measured sound speed in LS as a function of temperature for three measurement in the lab (Test 1, 2, 3). The dash line shows a linear fit to the reference data set, Test 3, and the shaded band represents the corresponding $\pm1\sigma$ residual spread about the fit, together with the propagated contribution from the temperature uncertainty of $0.1~^\circ\mathrm{C}$, giving a total width of $0.92~\mathrm{m/s}$. The other two data sets follow the same temperature dependence with few outlying points. The final result for the sound speed vs. temperature relation is: $ V_\mathrm{sound, LS} = 1469.69 - 3.916 \times T~(\mathrm{m/s})$ in the range of  $18.0\sim22.5~^\circ\mathrm{C}$.}
\label{fig4}
\end{figure}

Once calibrated, the dedicated setup was utilized to measure the sound speed directly within the LS. Repeated measurements (Test 1, 2, 3) show a stable and reproducible trend: the sound speed in LS decreases with increasing temperature, as shown in Fig.~\ref{fig4}. 

\begin{figure}[!hbt]
\centering
\includegraphics[width=0.6\linewidth]{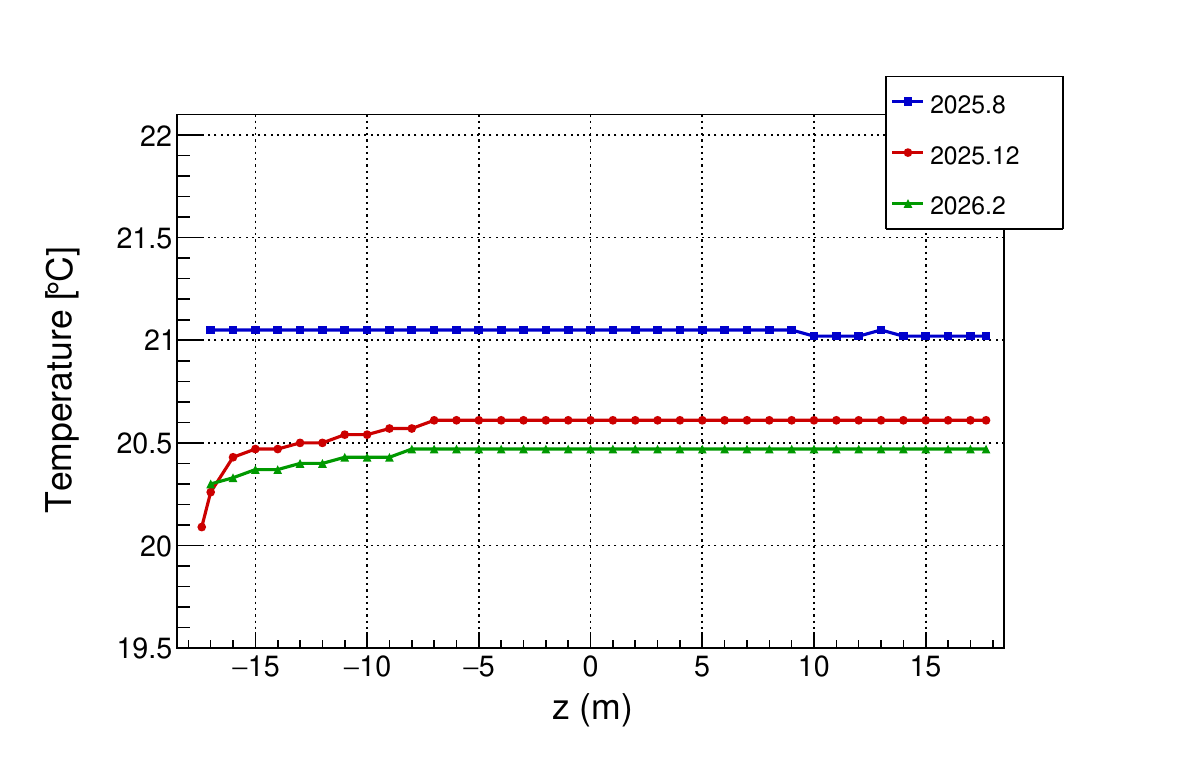}
\caption{Temperature profiles measured in the JUNO central detector during three data-taking campaigns. The temperature sensor was deployed with the ACU and recorded one measurement every 1 m along the central axis. The observed temperature change is primarily driven by the inherent difficulty in establishing perfect thermal equilibrium across the massive medium. In this analysis, the temperature field within the LS is assumed to be horizontally uniform, varying dominantly along the $z$-axis. The sound speed measurements are followed by every calibration campaign, and the values are used in the reconstruction was determined from the temperature at the corresponding source position.}\label{fig5}
\end{figure}

In JUNO, the temperature of the water pool is regulated by the circulation system, while the temperature inside the central detector is mainly governed by thermal diffusion through the acrylic. Before each calibration campaign, the temperature profile in the central detector was measured by deploying a temperature sensor with the ACU in the central axis, as shown in Fig.~\ref{fig5}. 

\subsection{Arrival Time Reconstruction Algorithm}
\label{sec::3.2}
\begin{figure}[!hbt]

\centering
\includegraphics[width=0.6\textwidth]{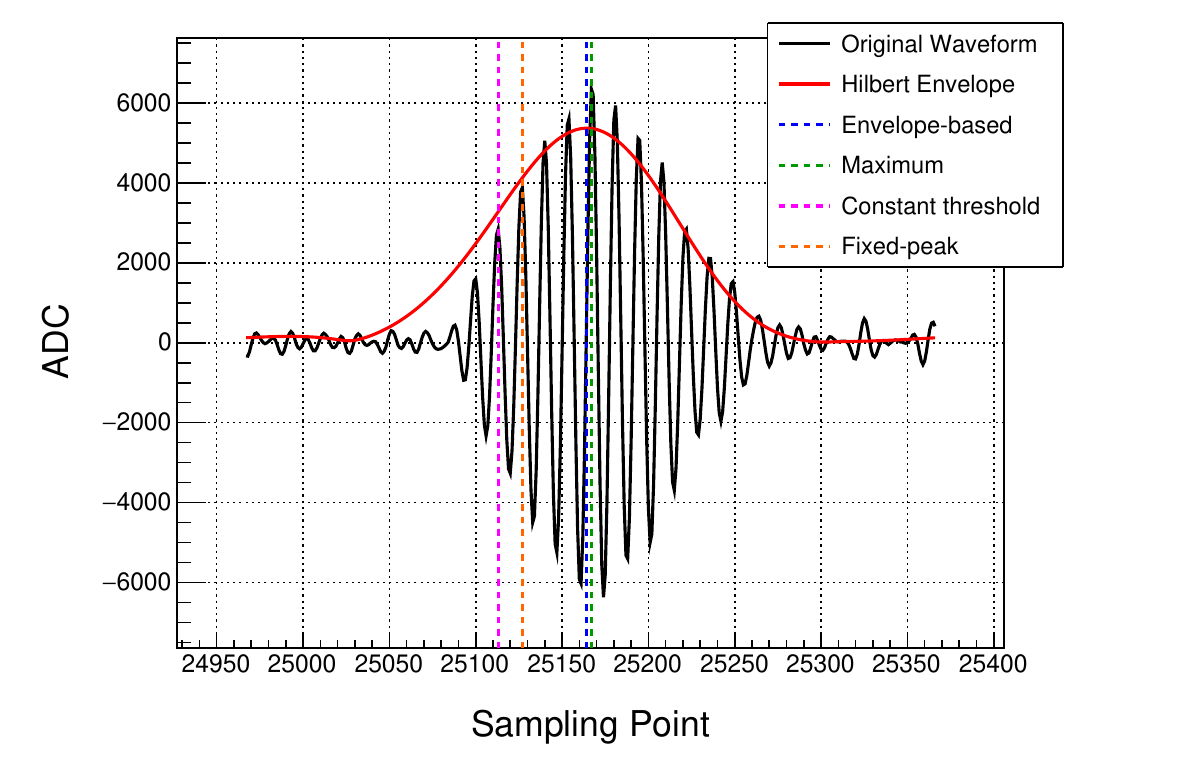}
\caption{Comparison of different arrival time extraction algorithms on a single-event waveform. Each sampling point represents 0.512~µs. The black solid line represents the pre-processed raw waveform, while the red solid line displays the amplitude envelope calculated via the Hilbert transform with a smooth process. The vertical dashed lines indicate the extracted timing positions using four different methods: envelope-based (blue), maximum (green), constant threshold (magenta), and fixed-peak (orange). The plot illustrates timing response and temporal offset of each algorithm relative to the pulse's rising edge.} \label{fig:wave}

\end{figure}

The acoustic signal propagation time, often referred to as the time-of-flight, is determined by the exact difference between the signal arrival time and the initial trigger time. In our system, the trigger time serves as the absolute time reference for the acoustic emission from the emitter control box~\cite{sym16091218}, whereas the arrival time must be accurately extracted from the received waveform. The trigger time is aligned to the emission time, while the arrival time needs to be read from the receiving waveforms. The recorded signal is not a single ideal pulse, but a finite-length acoustic burst consisting of multiple carrier cycles together with a damped trailing component, as shown in Fig.~\ref{fig:wave}. During transmission, the piezoelectric element is driven by an external tone burst to operate in a forced oscillation regime at its resonant frequency of 142~Hz. The cycle number of this burst is set to 7, which is optimized to have most stable control performance. Once the excitation ends, it transitions into free oscillation, functioning as an underdamped harmonic oscillator that generates an exponentially decaying ring-down signal. This waveform morphology makes the determination of the signal arrival time non-trivial. Several methods were therefore investigated for extracting a precise arrival time from the recorded waveform.

\begin{enumerate}
    \item \textbf{Maximum point method:} the arrival time is defined as the time of the global waveform maximum.

    \item \textbf{Constant-threshold method:} the arrival time is defined as the first time at which the waveform crosses a predefined threshold.

    \item \textbf{Envelope-based method:} a Hilbert transform is applied to obtain the signal envelope, and the arrival time is defined as the time at which the envelope reaches a fixed fraction of its maximum value.

    \item \textbf{Fixed-peak method:} the arrival time is defined as the time of a specified peak within the wave packet, identified according to the same inspection rules for all signals.
\end{enumerate}

\begin{figure}[!hbt]
    \centering
    \includegraphics[width=0.6\linewidth]{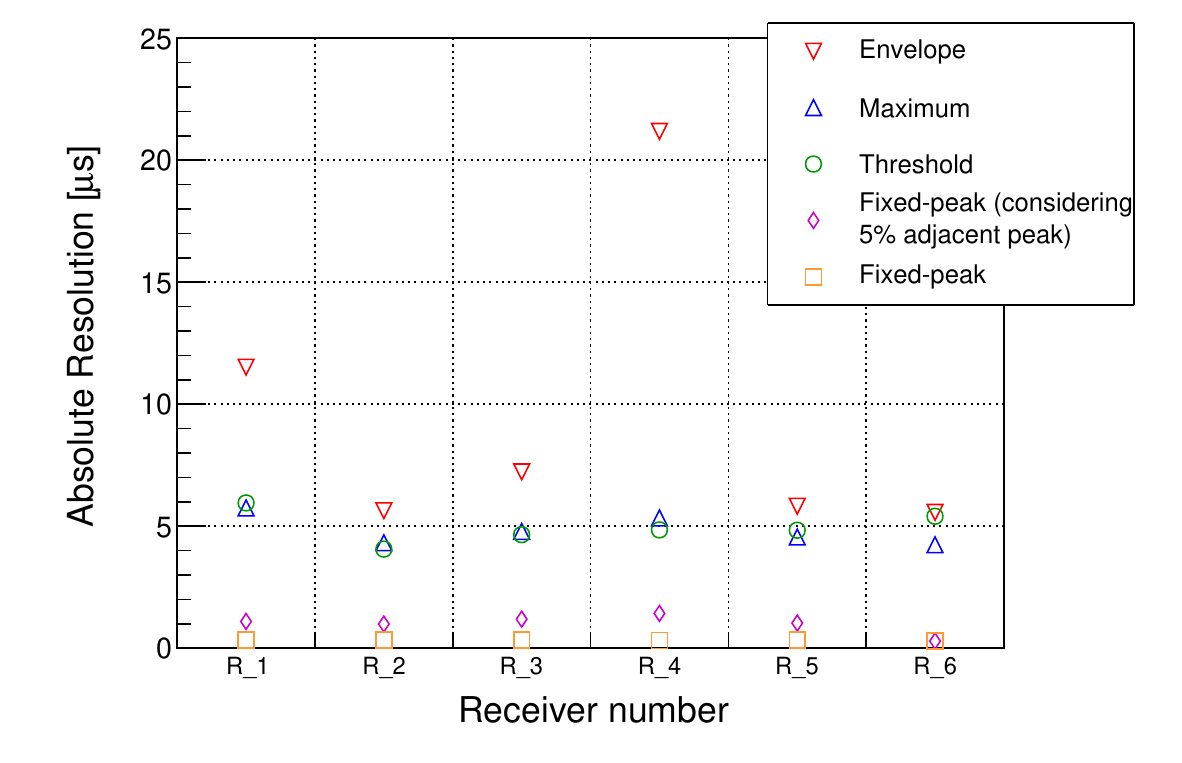}
    \caption{Absolute arrival-time resolution for the six active receivers obtained with four time-reconstruction methods, using the detector center as an example. The resolution is defined as the standard deviation of the reconstructed arrival-time distribution. An approximate 5\% probability of incorrectly selecting an adjacent peak is incorporated, which accounts for the fraction of waveforms severely degraded by noise. The envelope-based method gives the poorest performance, while the threshold and maximum point methods are comparable and the fixed-peak method provides the best overall stability.}
    \label{fig:4tof}
\end{figure}

The performance of the different arrival time reconstruction methods was evaluated using real waveforms measured in JUNO, as shown in Fig.~\ref{fig:4tof}. Based on this study, the fixed peak method was selected for the final offline analysis because it provides the best timing resolution for the present application, while the maximum point method is chosen in the online position monitoring.

\subsection{Position Reconstruction}

The source position reconstruction is formulated as a non-linear optimization problem based on the time-of-flight information from six receivers. All receivers are assigned equal weights, since their responses are stable and no significant outliers are observed. Although the quadratic range equations can be transformed into a linear system to solve, the geometric degeneracy of the present receiver configuration, especially in the $z$ direction, makes the linearized solution less well constrained and therefore less reliable. We therefore use the \textbf{TMinuit} \texttt{Migrad} minimizer~\cite{James1975MINUIT} for the final reconstruction. The fit minimizes the sum of squared residuals between the measured and predicted time-of-flight, yielding the optimal source coordinates, see eq.~\eqref{eq}. 

\begin{equation}
\label{eq}
S(\mathbf{X}, t_0) = \sum_{i=1}^{N}
\left[
t_i^{\mathrm{meas}} - \left( t_0 + \frac{\left| \mathbf{X} - \mathbf{R}_i \right|}{v_s} \right)
\right]^2
\end{equation}
where \(t_i^{\mathrm{meas}}\) is the measured arrival time at the \(i\)-th receiver, \(t_0\) is the emission time, \(\mathbf{X}=(x,y,z)\) is the source position to be reconstructed, \(\mathbf{R}_i=(x_i,y_i,z_i)\) is the position of the \(i\)-th receiver, \(v_s\) is the speed of sound in the LS, and \(N\) is the number of active receivers. The linearized solution provides the initial seeds for the optimization, while the path integral is performed with a constant bin width of $0.01\,\mathrm{m}$ for temperature changes. Compared with the linearized solution, the non-linear fit fully exploits the timing information, improves the robustness of the reconstruction in the weakly constrained $z$ direction, and provides parameter uncertainties through the covariance matrix at the minimum.

For online monitoring, the arrival time is extracted from the waveform maximum after applying a timing-offset correction relative to the burst start, and the source position is reconstructed directly with \textbf{TMinuit} \texttt{Migrad}. For the final offline reconstruction, the arrival time is determined from the fixed-peak method, followed by the same timing-offset correction. An initial seed position $\mathbf{X}_0$ is obtained from a linear least-squares solution and then refined with \textbf{TMinuit} \texttt{Migrad}.

\section{USS calibration}
\label{sec:calibration}

To account for possible drifts in the receiver positions, an in-situ calibration of the USS receiver geometry is performed before the CLS calibration campaign. An ultrasonic emitter is deployed along the central axis with the ACU which served in Daya Bay~\cite{DayaBayACU2016}, so that its position is accurately controlled and known from the deployment length. By scanning multiple positions along the axis and extracting the time-of-flight of the signals to all available receivers, the effective receiver geometry under detector operating conditions can be determined. 

The calibration is formulated as a global fit to the measured time-of-flight. The expected time-of-flight is calculated from the emitter-receiver geometric distance, assuming a constant sound speed within the temperature-stable region in which this calibration was conducted. The dependence of the speed of sound on temperature is discussed in Sec. 3.1. The receiver coordinates are treated as free parameters, whereas the emitter positions are fixed at the pre-defined coordinates in the PMT reference frame. The surveyed receiver positions obtained before LS filling are used as the initial values in the fit.

This multi-parameter and highly correlated nonlinear optimization problem is solved with the Ceres Solver~\cite{AgarwalCeresSolver} framework by minimizing the residuals between the measured and predicted time-of-flight for all calibration points simultaneously. The procedure yields the effective receiver coordinates under filled-detector conditions, revealing structural displacements of $1-10\,\mathrm{cm}$ relative to the initial dry-survey positions due to survey imperfections and acrylic deformation. Furthermore, the fitting algorithm inherently evaluates the parameter covariance, yielding a statistical uncertainty of $0.82\,\mathrm{cm}$ for these newly calibrated coordinates. The calibrated receiver geometry is then used as the input for the USS-based position reconstruction in the CLS calibration campaign described in the next section.

\begin{figure}[hbt!]
    \centering
    \begin{subfigure}{0.42\textwidth}
        \centering
        \includegraphics[width=1\linewidth]{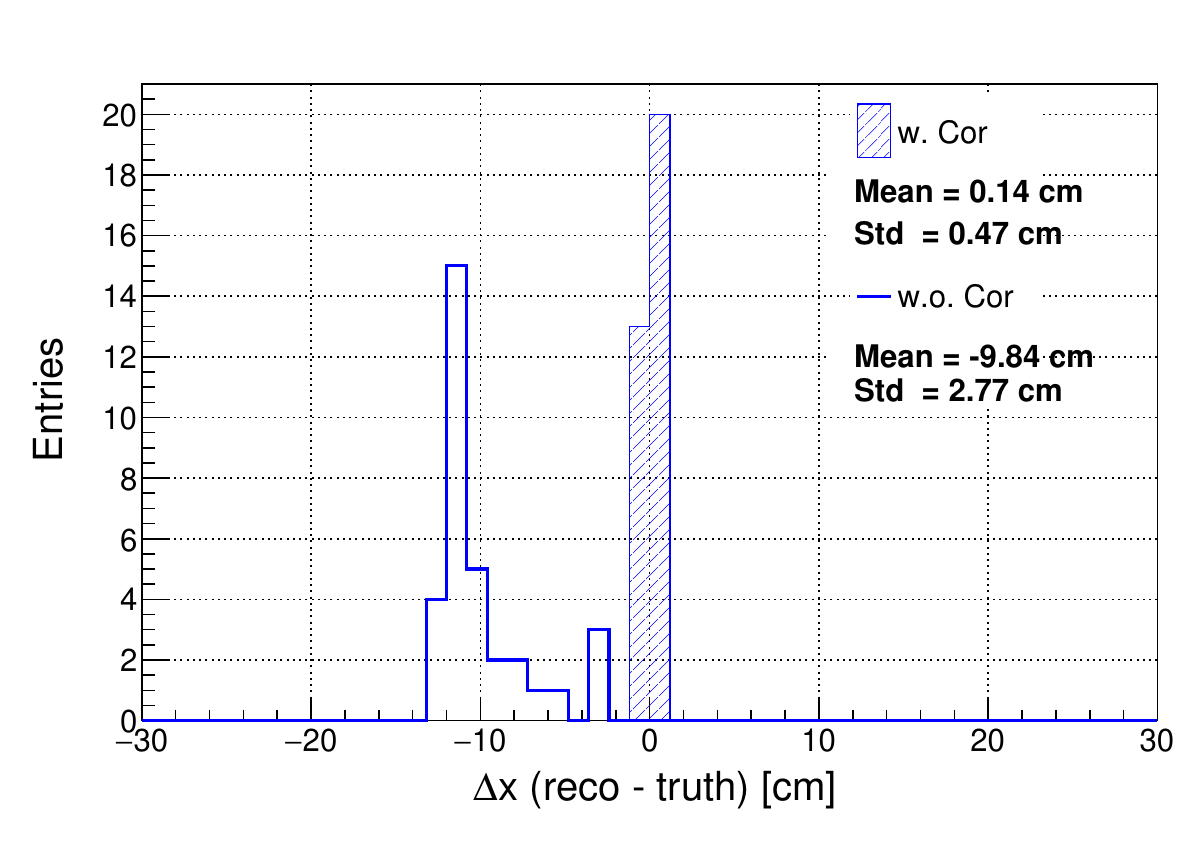}

    \end{subfigure}
    \begin{subfigure}{0.42\textwidth}
        \centering
        \includegraphics[width=1\linewidth]{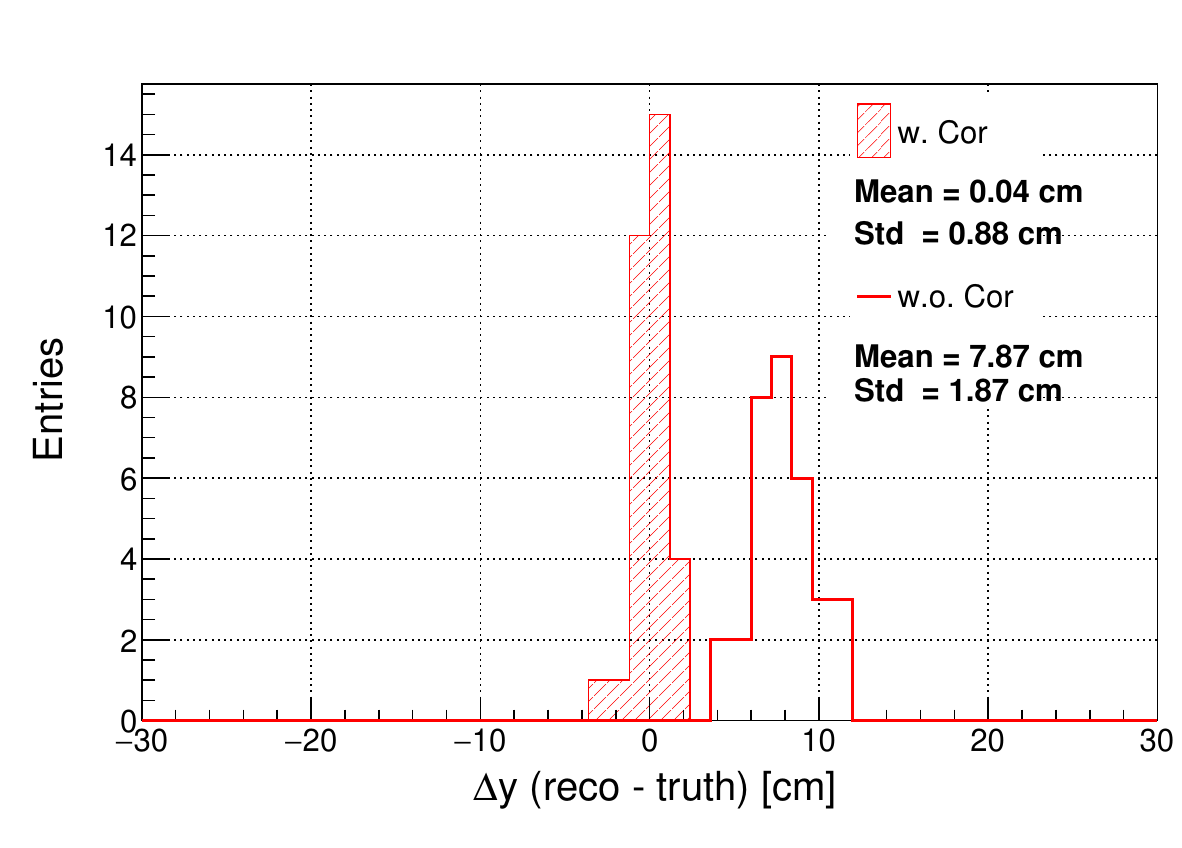}
        
    \end{subfigure}
        \begin{subfigure}{0.42\textwidth}
        \centering
        \includegraphics[width=1\linewidth]{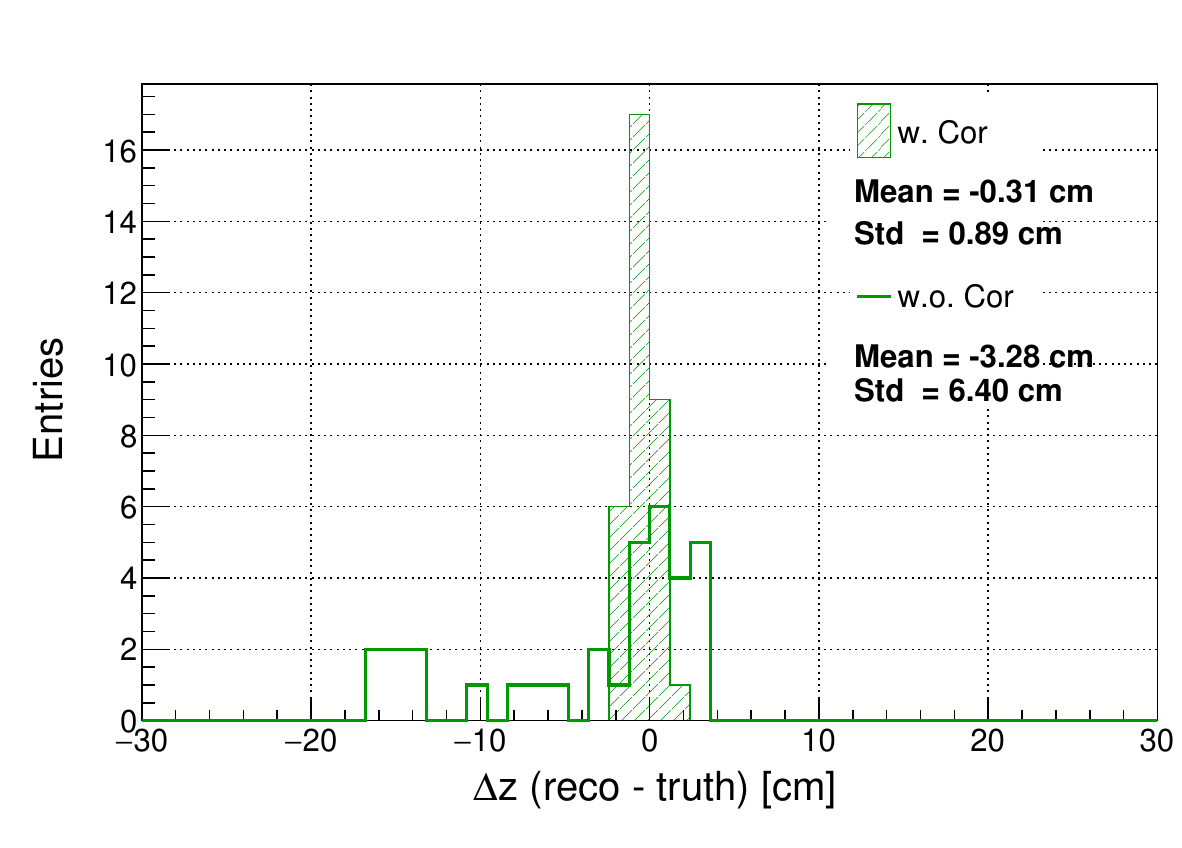}

    \end{subfigure}
        \begin{subfigure}{0.42\textwidth}
        \centering
        \includegraphics[width=1\linewidth]{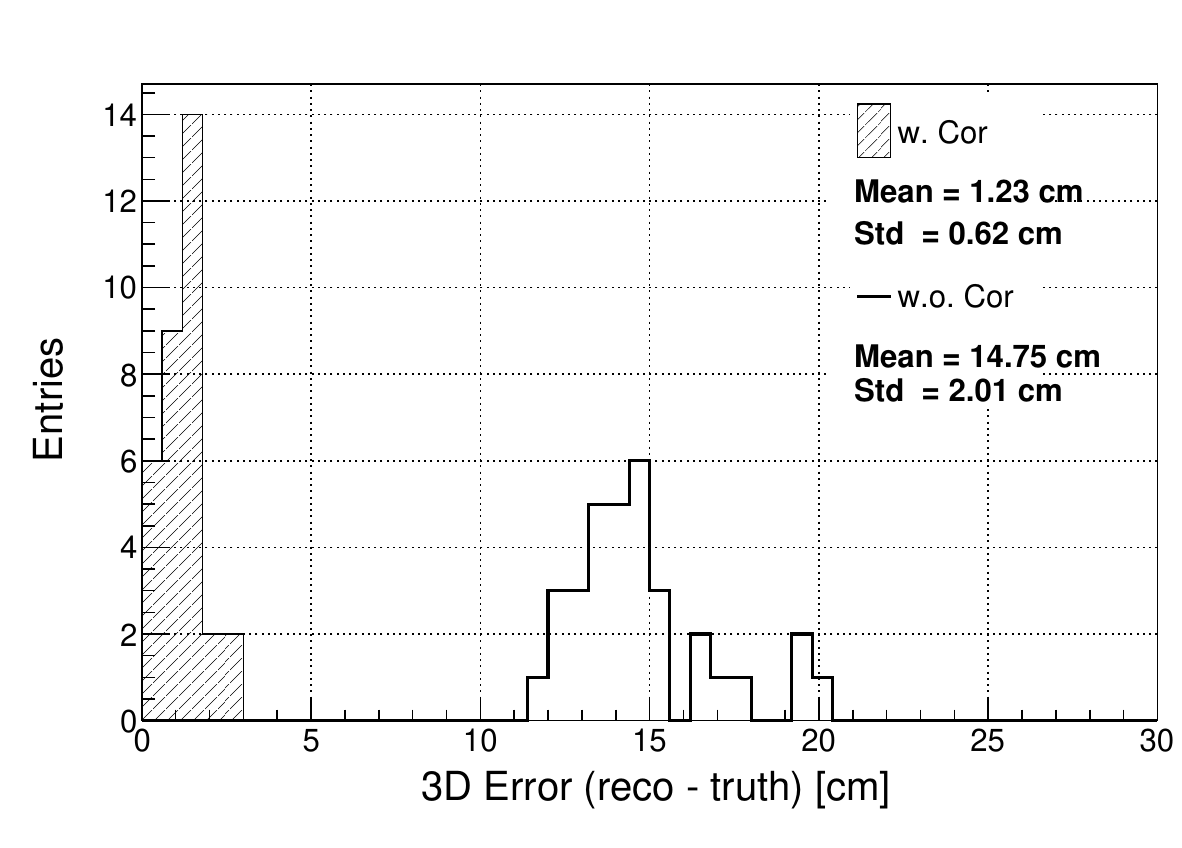}
        
    \end{subfigure}
    \caption{Residual distributions of the reconstructed x, y, z coordinates and the overall 3D error before and after the in-situ calibration of the USS receiver positions. During the calibration process, a single emitter is positioned at 33 predefined positions along the central axis. The residuals are defined as the differences between the USS-reconstructed coordinates and the predefined positions. Open histograms use the receiver coordinates from the pre-filling survey (w.o. Cor), while shaded histograms use the calibrated coordinates (w. Cor). The reduced offsets and widths after calibration indicate improved reconstruction performance.}
    \label{fig:USScalibrationResult}
\end{figure}

In the present analysis, only six receivers remain active and their effective positions were obtained. As a consequence, the effective receiver geometry is less favourable than originally designed and provides weaker geometrical constraints, especially in the vertical direction. Despite this mathematically less-constrained configuration, the reconstruction accuracy along the central axis is significantly improved after the calibration. This accuracy is quantified by the mean positioning error—defined as the spatial distance between the known deployment position and the reconstructed coordinates (Fig.~\ref{fig:USScalibrationResult}). The overall mean positioning error is $1.23~\mathrm{cm}$.

\section{Calibration Campaign in the JUNO Central Detector}

\subsection{Online Monitoring and Waveform Accumulation}

During the deployment of the calibration source within the central detector, operators lack precise, real-time spatial information, relying instead on approximate coordinates inferred from the two stainless-steel control cables. Therefore, continuous online monitoring is strictly required to navigate the source safely away from acrylic boundaries, to transition accurately between calibration points, and to guarantee sufficient spatial coverage of the planned scan region. Furthermore, precise spatial feedback is required by the JUNO Online Event Classification System~\cite{Fang2026OECSWJUNO}. Although the OEC is primarily designed for general online event selection to process the overwhelming raw data volume, its specialized filtering strategy during calibration campaigns facilitates the selection of genuine calibration events, which are highly localized around the source.

Because the maximum point method is a robust and computationally fast timing reconstruction method, a time-of-flight estimate can be obtained immediately after the waveform is digitized. For each calibration point, the full online reconstruction takes about 30 s, including repeated ultrasonic emissions, waveform conversion, time-of-flight extraction, and position reconstruction. The reconstructed position is then provided to the JUNO online event-classification system, which selects a spherical region with a radius of 4 m around the estimated source position. Although the online time resolution is worse than that of the offline reconstruction (see Fig.~\ref{fig:4tof}), it is sufficient for this purpose.

For the final offline reconstruction, USS data are accumulated while the calibration source is held at a fixed position. The ultrasonic pulse is emitted once per second to collect sufficient waveforms. In a typical data set, 100 waveforms are selected for the accurate positioning of one calibration point, ensuring that random interference is excluded.

\subsection{Offline USS Reconstruction}
\begin{figure}[!hbt]
    \centering
    \includegraphics[width=0.7\linewidth]{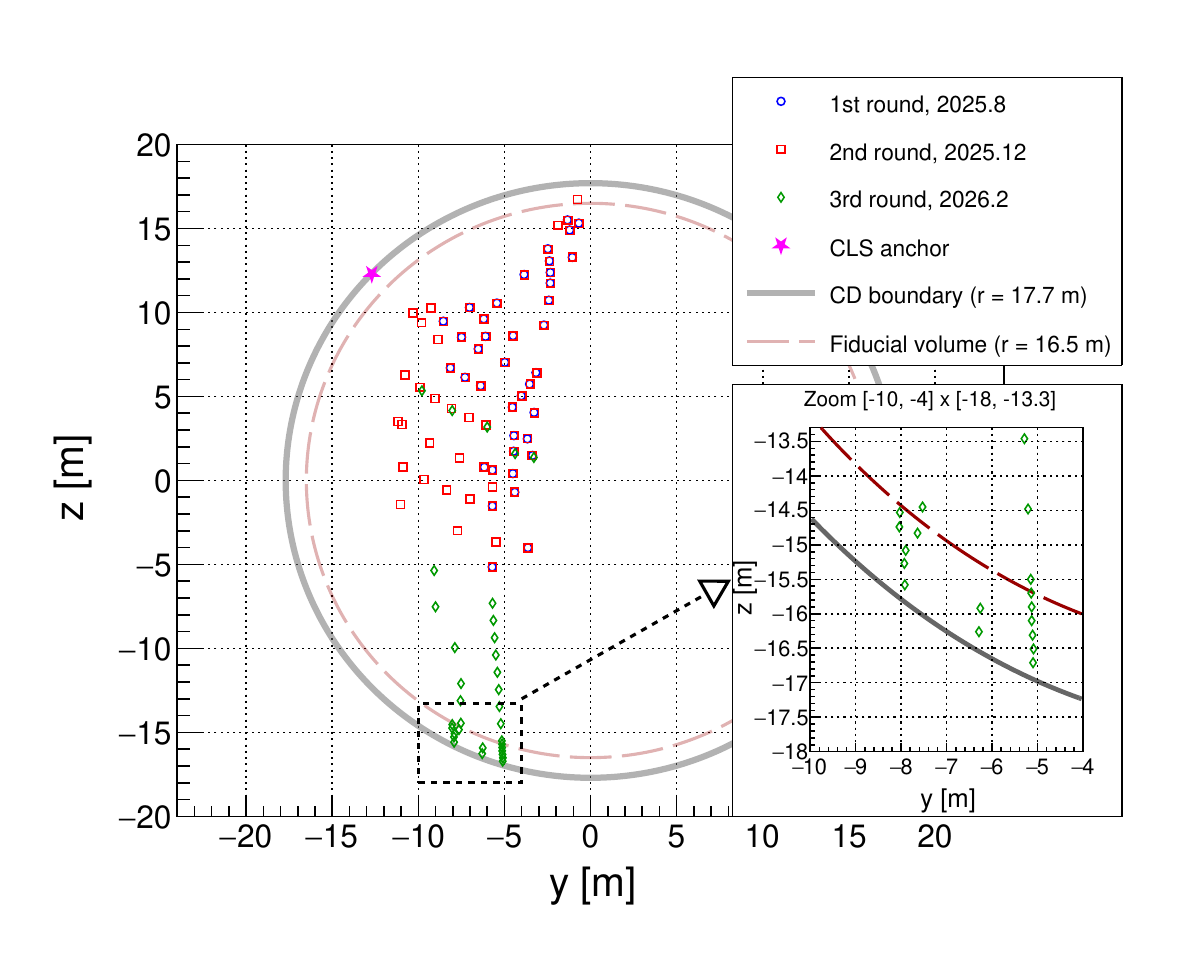}
    \caption{Distribution of the reconstructed CLS calibration points in the $y$-$z$ plane across the three calibration campaigns. Different markers distinguish the respective campaigns, while the magenta star and the solid circle denote the CLS anchor and the Central Detector (CD) boundary ($R = 17.7~\mathrm{m}$), respectively. The reconstructed points effectively cover the accessible off-axis scan region. The bottom-right inset provides a magnified view of the positions located outside the fiducial volume (FV), demonstrating the system's high positioning reliability and its robustness against complex boundary effects, such as acoustic reflections.}
    \label{fig:offline}
\end{figure}

The offline USS reconstruction aims to achieve the highest possible positioning accuracy. For each calibration point, the mean value of the reconstructed coordinates from the corresponding 100 waveforms is taken as the final source position. Three CLS calibration campaigns, containing 40, 63, and 34 calibration points, respectively, were performed to construct the detector non-uniformity map, as shown in Fig.~\ref{fig:offline}. 

\section{Positioning Performance Analysis}
\label{sec:Uncertainty}
The ultimate positioning performance of the ultrasonic system is governed by the precision of both the timing extraction and the position reconstruction algorithms. This section first evaluates the physical and instrumental sources of uncertainty, derives their representative magnitudes directly from experimental data, and subsequently utilizes a dedicated simulation framework to quantitatively assess their combined impact on the final source coordinates.

The main sources of uncertainty is as follows:
\begin{enumerate}
    \item \textbf{Arrival-time resolution:} The reconstructed arrival times are smeared to account for both instrumental limitations and algorithmic timing fluctuations. Building upon the algorithm evaluation in Sec.~\ref{sec::3.2}, this analysis focuses on the physical and hardware-induced timing uncertainties. The timing spread inherently arises from several factors: variations in the effective acoustic emission and reception centers, digitizer sampling jitter, piezoelectric vibration inconsistencies, and minor emission trigger delays. Under actual JUNO operating conditions, macroscopic effects such as local sound-speed variations and source pendulum motions could also contribute. To isolate these influences, the measured time-of-flight resolution is analyzed across a wide range of propagation distances, as shown in Fig.~\ref{fig:placeholder}. The resolution remains remarkably stable at approximately $0.5~\mu\mathrm{s}$. Given that the digitizer's sampling interval is $0.512~\mu\mathrm{s}$, this distance-independent plateau confirms that the overall timing spread is strictly dominated by the intrinsic limitations of the front-end electronics and transducers, effectively representing the hardware limit.
\begin{figure}[!hbt]
    \centering
    \includegraphics[width=0.6\linewidth]{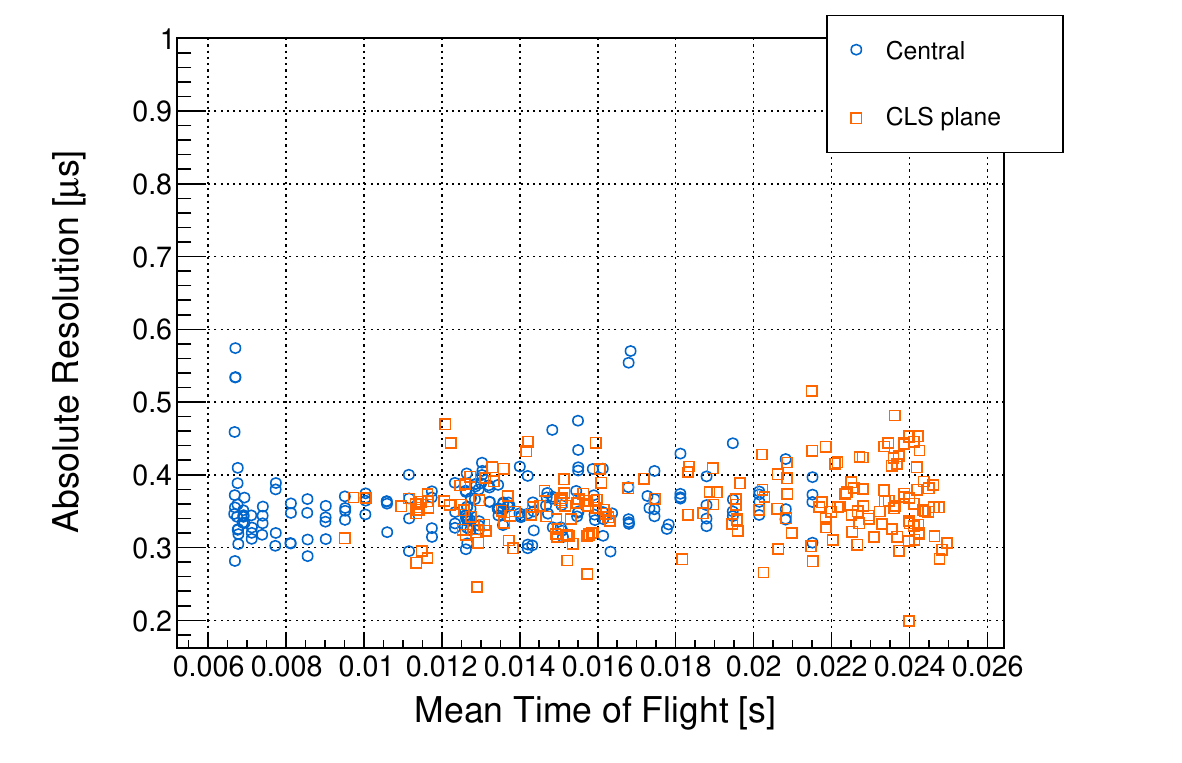}
    \caption{Absolute arrival-time resolution as a function of mean time-of-flight for different calibration scenarios, including the ACU-based USS calibration and the CLS calibration. In all cases, the arrival time is reconstructed with the same fixed-peak method. An adjacent peak assignment would shift the reconstructed time by 7~µs. The outliers are associated with periods of abnormally strong electrical noise in one receiver, which leads to larger timing fluctuations.}
    \label{fig:placeholder}
\end{figure}
    \item \textbf{Reference-frame uncertainty:} The USS reference frame is nominally assumed to coincide with the global PMT reference frame. The receiver coordinates utilized in the reconstruction are derived from the in-situ USS calibration, yielding a typical fitting uncertainty of approximately $0.8~\mathrm{cm}$. To evaluate the propagation of this error, the receiver coordinates in the simulation are independently smeared using Gaussian distributions with standard deviations of $0.8~\mathrm{cm}$, and the subsequent impact on the reconstructed source position is quantified.

    \item \textbf{Adjacent-peak misidentification:} During the fixed-peak arrival-time extraction, there is a small probability—empirically modeled at 5\% based on the manual counting of ambiguous waveforms—of missing the correct acoustic peak. An incorrect peak assignment shifts the reconstructed time-of-flight by exactly one wavelength period of the transducer's central frequency, corresponding to a $7~\mu\mathrm{s}$ offset. To rigorously reflect this discrete systematic deviation in our simulation framework, we assume that random electronics noise degrades the rising and falling edges of the signal envelope symmetrically. Therefore, this effect is conservatively implemented in the simulation as an equal 2.5\% chance for the time-of-flight to be erroneously pulled early (a pre-trigger shift of $-7~\mu\mathrm{s}$) and a 2.5\% chance to be triggered late (a post-trigger shift of $+7~\mu\mathrm{s}$) for any given acoustic evaluation.
\end{enumerate}

\subsection{Simulation Chain}
\begin{figure}[!hb]
    \centering
    \includegraphics[width=0.7\linewidth]{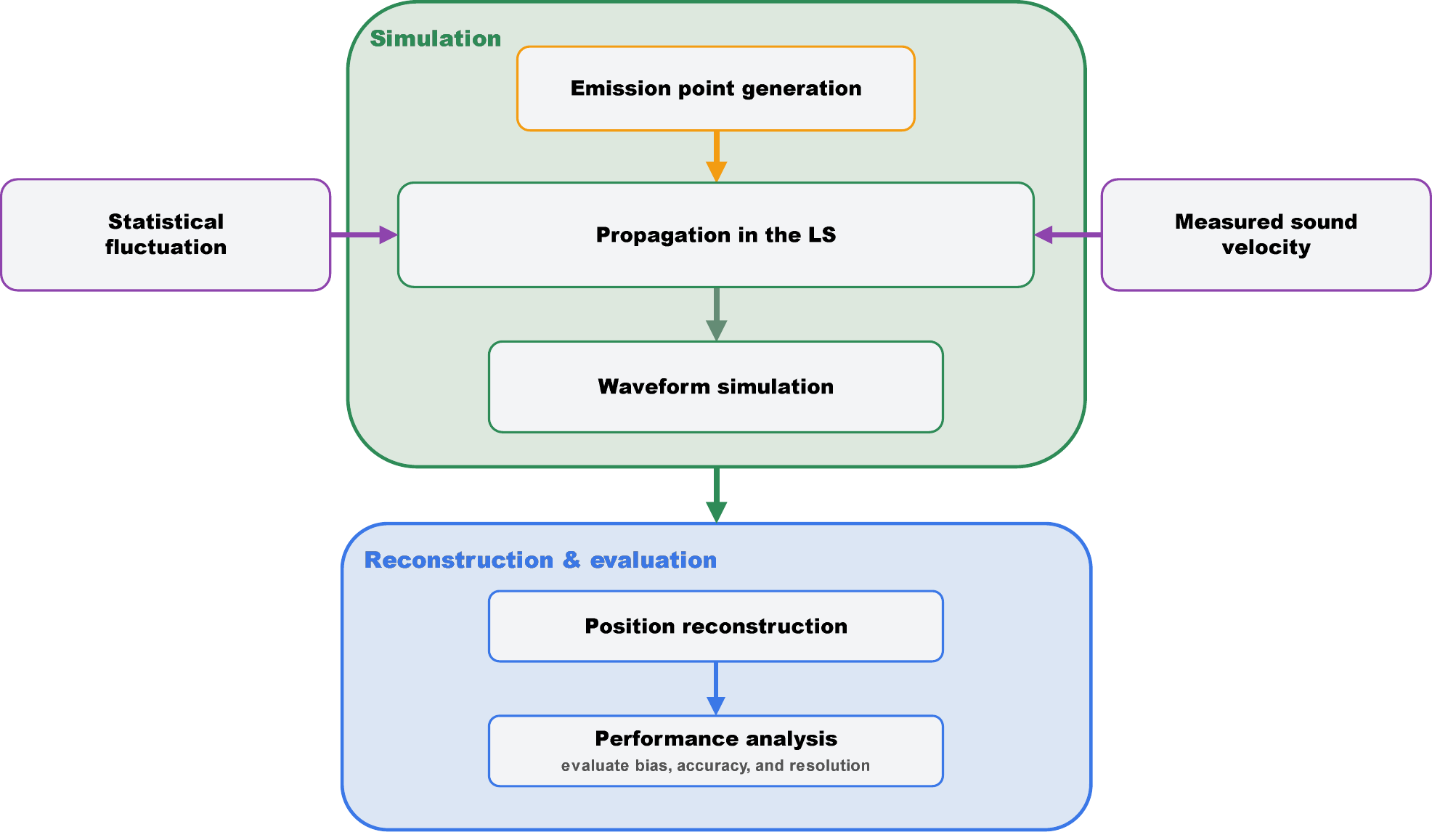}
    \caption{Simulation workflow used to evaluate the performance of the acoustic positioning system. The chain includes (1) emission-point generation for calibration scans, (2) sound propagation in LS with the temperature-dependent sound-speed model, (3) waveform generation with noise and digitization, (4) position reconstruction from time-of-flight information, and (5) performance evaluation.}\label{fig:simulationflow}
\end{figure}
\begin{figure}[!t]
    \centering
    \begin{subfigure}{0.45\textwidth}
        \centering
        \includegraphics[width=1\linewidth]{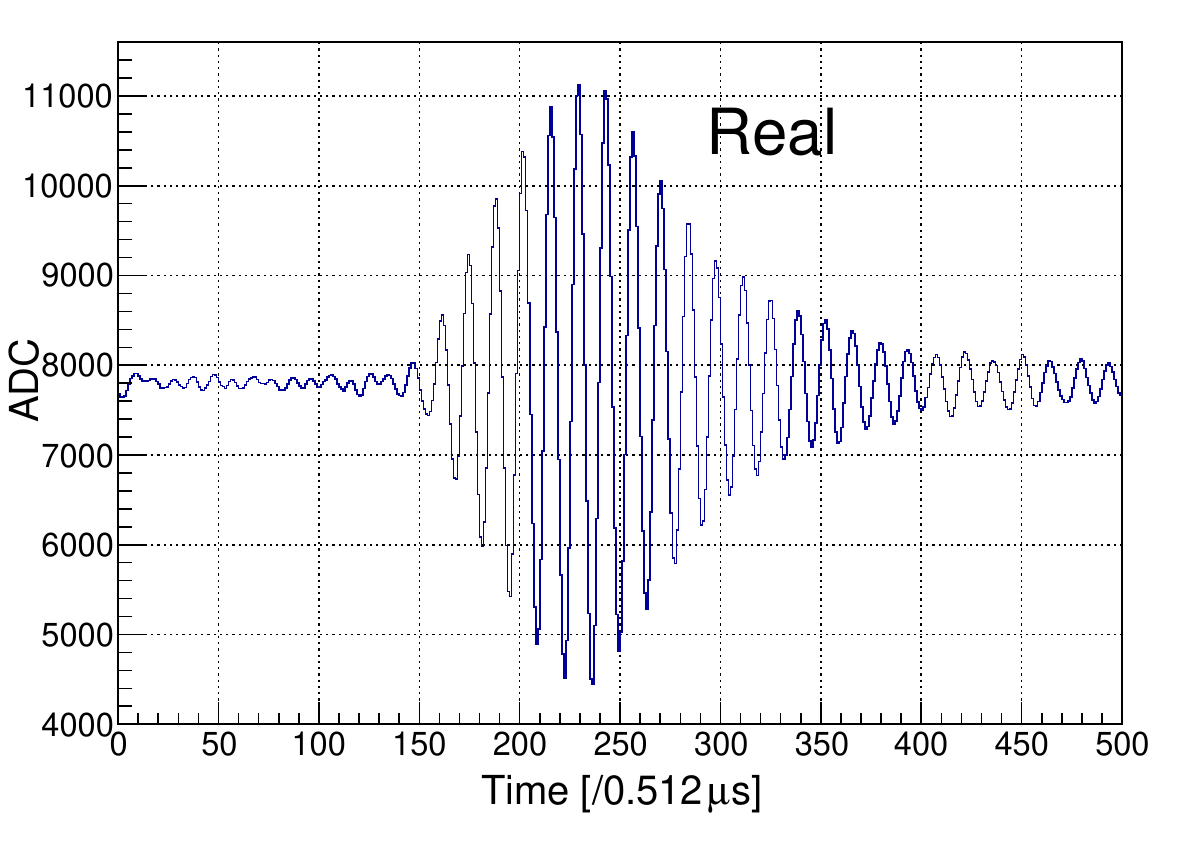}
        \caption{}
        \label{fig:subfig1}
    \end{subfigure}
    \hfill
    \begin{subfigure}{0.45\textwidth}
        \centering
        \includegraphics[width=1\linewidth]{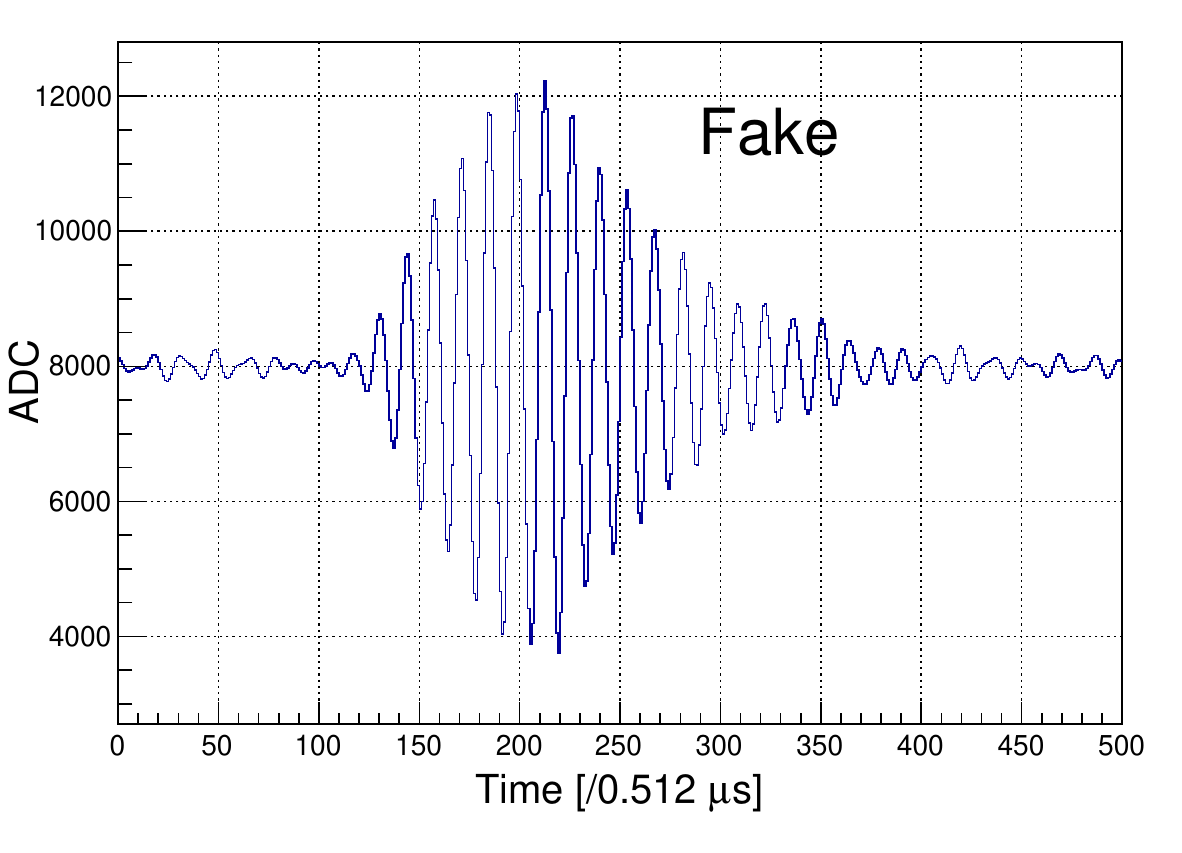}
        \caption{}
        \label{fig:subfig2}
    \end{subfigure}
    \caption{Representative real and simulated waveforms: (a) a waveform measured in JUNO and (b) a waveform generated in the simulation. In both cases, the signal amplitude rises over the first several oscillation cycles, reaches a maximum near the last emitted cycle, and is followed by a damped tail. In the simulation, the leading envelope is modelled with a parabolic rise and the trailing part with an exponential decay.}
    \label{fig:combined}
\end{figure}

A dedicated simulation was developed to evaluate the performance of the final position reconstruction algorithm, as illustrated in Fig.~\ref{fig:simulationflow}. To assess how the intrinsic time resolution affects the ultimate spatial reconstruction, the simulated arrival times are first subjected to a smearing process before being processed through the positioning algorithm. The characteristics of the real waveforms are mimicked to build analytical fake waveform, as shown in Fig.~\ref{fig:combined}. Each wave envelope, representing a single emission, contains several pulses. The sampling period is 0.512 $\mu$s, so that one wave envelope contains approximately 100 sampling points. The nominal signal-to-noise ratio (SNR), accounting for both waveform and baseline fluctuations, is approximately 17. Other parameters were chosen together to reproduce the experimental conditions expected in the JUNO detector, summarized in Table~\ref{tab:sim_params}. The main uncertainty sources are introduced in a unified way, including waveform noise, arrival time fluctuations, sound speed variation, incorrect peak identification, and the uncertainty of the calibrated receiver positions. To simulate the positioning procedure for the calibration, 100,000 emission points were uniformly generated within the CLS plane. The time of flight to each of the 6 receivers was calculated for each point, incorporating the speed of sound model described above. For each simulated event, waveforms are generated for each receiver, respectively, with their fourth peak aligned to the calculated arrival time. Gaussian timing jitter with a width of 0.5~µs was added to each arrival time to model time resolution.\\

\begin{table}[!ht]
\centering
\small
\setlength{\tabcolsep}{6pt} 
\renewcommand{\arraystretch}{1.2} 
\begin{tabular}{c c}
\toprule
\textbf{Simulation Parameter} & \textbf{Value / Range} \\
\midrule
Signal-to-noise ratio (Amplitude) & 17 \\
Cycles number (in Sec.~\ref{sec::3.2}) & 7 \\
Speed of sound & $1387.26 \pm 0.9\,\mathrm{m/s}$ \\
Time smearing & $\sigma_{t} = 0.5\,\mu\mathrm{s}$ \\
Receiver position uncertainty & $\sigma = 0.8\,\mathrm{cm}$ \\
Adjacent peak probability & \makecell[l]{5\% with a time \\shift of $\pm 7\,\mu\mathrm{s}$} \\
\bottomrule

\end{tabular}
\caption{Simulation parameters used in the positioning performance analysis.}
\label{tab:sim_params}
\end{table}

\begin{figure*}[!ht]
    \centering
    \begin{subfigure}{0.8\textwidth}
        \centering
        \includegraphics[width=1\linewidth]{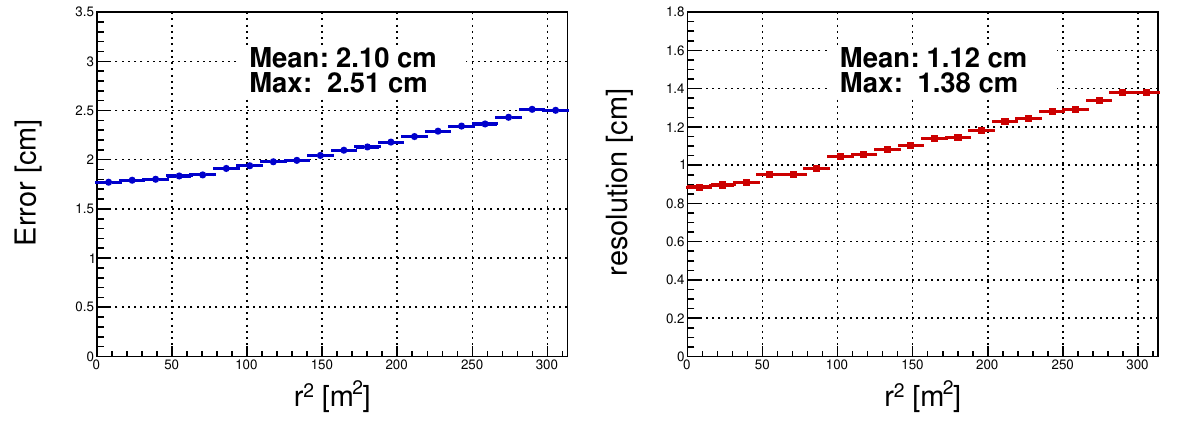}

        \label{fig:subfig1}
    \end{subfigure}

    \vspace{0.5cm}
    
    \begin{subfigure}{0.6\textwidth}
        \centering
        \includegraphics[width=1\linewidth]{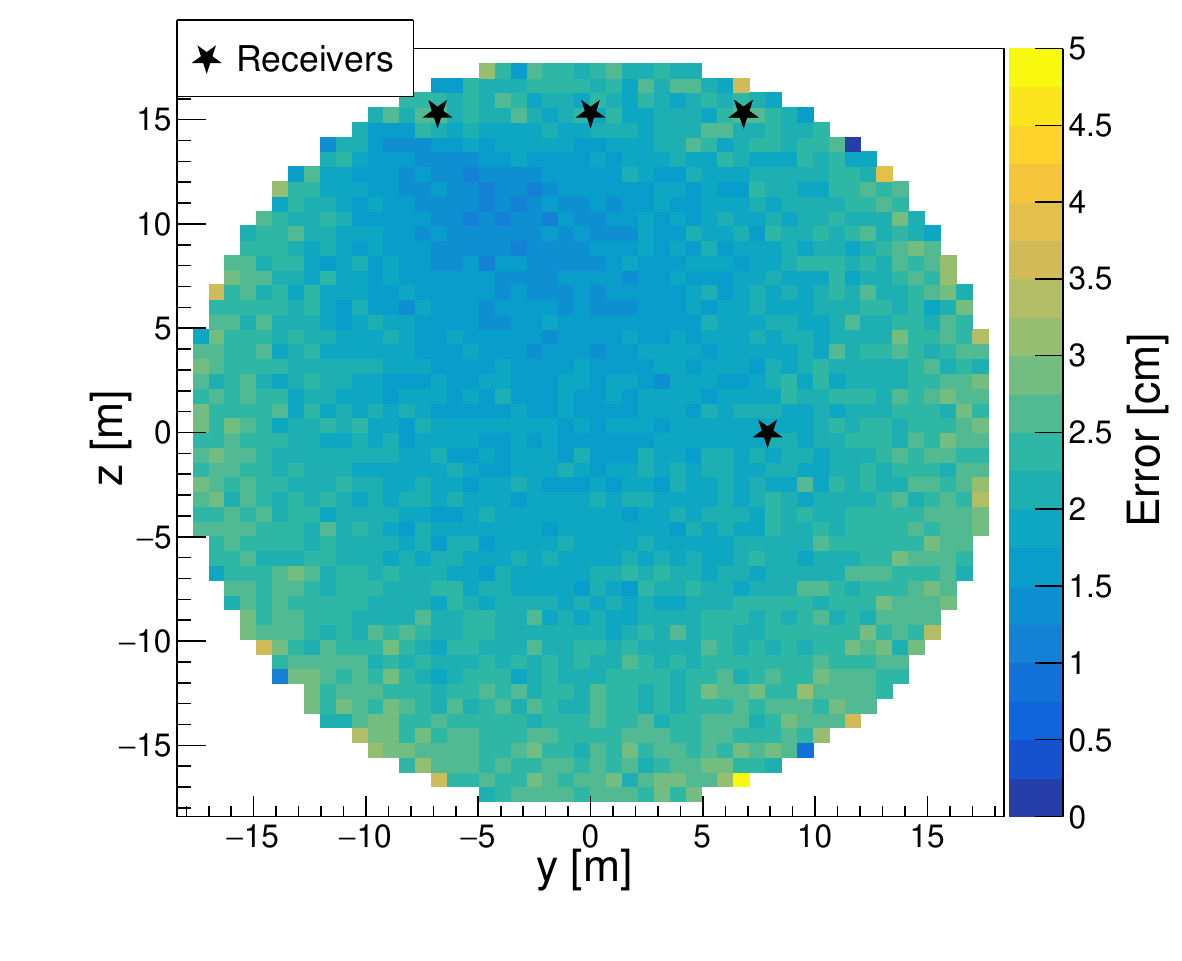}  

        \label{fig:subfig4}
    \end{subfigure}
    
    \caption{Performance of the USS reconstruction in the CLS plane for 100000 uniformly generated source positions. The mean position error and the reconstruction resolution as functions of the true $r^2$ are shown in the top-left and top-right panels, respectively. The bottom panel shows the two-dimensional map of the mean position error in the y-z plane, where the locations of the USS receivers are indicated by star markers. The error is defined as the mean distance between the reconstructed and true positions, and the resolution is the standard deviation of the error distribution within each bin.}
    \label{fig:bias}
\end{figure*}
The same reconstruction procedure is then applied to the simulated USS data. The performance is quantified by the positioning error, defined as the mean distance between the reconstructed position and the true source position, and resolution, defined as the standard deviation of the position-error distribution. The resulting error distributions are shown in Fig.~\ref{fig:bias}. Including all effects listed in Table~\ref{tab:sim_params}, the simulation gives a mean absolute position error below 2 cm and a reconstruction resolution better than 1 cm throughout the CLS plane. The spatial variation of the performance is mainly governed by the angular coverage of the receiver array. Because the active receivers are concentrated in the upper half, source positions viewed under a broader spread of receiver directions are reconstructed more accurately, while positions with poorer angular coverage show larger mean errors and worse resolutions. This behavior accounts for both the degradation in the lower region and the moderate deterioration near the acrylic boundary. The overall system performance is characterized by the root mean square error (RMSE), which combines the mean error and the resolution, yielding a value of 2.40 cm.\\

\section{Summary}
 
This work presents an ultrasonic positioning system developed for precise calibration source deployment in JUNO. By combining a sound-speed model, arrival time reconstruction, and position fitting, the system determines the three-dimensional coordinates of the source. In data, the central-axis deployment yields a mean error of 1.23 cm. A dedicated performance study that includes sound-speed variation, timing resolution, peak misidentification, and receiver-position uncertainty indicates a mean absolute position error of approximately 2~cm and a resolution of approximately 1~cm for the simulated reconstruction, and an overall uncertainty of 2.40~cm. These results establish the present system as a practical solution for centimetre-level source positioning in large liquid scintillator detectors and define clear directions for further improvement through geometry optimization and receiver redundancy.

\section{Discussion}

To place the performance of the present system in context, we compare published quoted absolute positioning uncertainties and their corresponding detector radii, as summarized in Table~\ref{tab:coordinates}. These figures show that centimetre-scale absolute positioning is already at the limit of large-detector mechanical metrology. Under the common $\Delta R/R$ metric, the achieved JUNO performance is comparable to the best published benchmarks, while additionally enabling off-axis deployment. 

\begin{table*}[!h]
\centering
\small
\setlength{\tabcolsep}{4pt} 
\renewcommand{\arraystretch}{1.25}

\begin{tabularx}{\textwidth}{c >{\centering\arraybackslash}X >{\centering\arraybackslash}X c c}
\toprule
\textbf{Experiment} & \textbf{Deployment mode} & \textbf{Reference scale} & \makecell[c]{\textbf{Quoted absolute}\\\textbf{uncertainty}} & \makecell[c]{\textbf{Relative}\\\textbf{uncertainty}} \\
\midrule
Daya Bay~\cite{DayaBayACU2016}
& Central-axis/ Radioactive source
& OAV radius $R=2.0\,\mathrm{m}$, $H=4.0\,\mathrm{m}$
& $\pm 0.4\,\mathrm{cm}$
& 0.20\% \\

Super-Kamiokande~\cite{ABE2014253}
& Central-axis/ light source
& ID radius $R=16.9\,\mathrm{m}$, $H=36.2\,\mathrm{m}$
& $\pm 1.0\,\mathrm{cm}$
& 0.06\% \\

KamLAND~\cite{article1}
& Off-axis pole/ Radioactive source
& LS balloon radius $R=6.5\,\mathrm{m}$
& $\pm 2.0\,\mathrm{cm}$
& 0.30\% \\

SNO+~\cite{Bayes_2020}
& Off-axis rope/ Radioactive source
& AV radius $R=6.0\,\mathrm{m}$
& $\pm 1.5\,\mathrm{cm}$
& 0.25\% \\

\textbf{JUNO (this work)}
& Off-axis CLS/ Radioactive source
& AV radius $R=17.7\,\mathrm{m}$
& $\mathbf{\pm 2.4\,\mathrm{cm}}$
& \textbf{0.14\%} \\
\bottomrule
\end{tabularx}
\caption{Illustrative comparison of quoted source-positioning or positioning-related uncertainties in large neutrino detectors. The relative values are estimated here as $\Delta R/R$ using the detector or active-volume radii listed in the third column. Because the quoted quantities are not strictly identical across experiments, the comparison is intended only to provide order-of-magnitude context.}
\label{tab:coordinates}
\end{table*}

Further improvements to the USS can be pursued through optimization of the receiver geometry. The source-position reconstruction is based on solving a nonlinear system constrained by time-of-flight measurements, and its stability depends strongly on the spatial distribution of the receivers. In the current configuration, partial geometrical degeneracies remain, leading to reduced sensitivity along certain directions. A more uniform and fully three-dimensional receiver arrangement would improve the conditioning of the reconstruction problem and enhance the stability of the solution. In particular, increasing the angular coverage of the receiver array would help reduce directional error and improve the resolution along weakly constrained axes.\\

\section*{Acknowledgements}

We gratefully acknowledge the support and collaborative efforts of the JUNO collaboration. This work is supported by the National Key Research and Development Program of China (Grant no. 2023YFA1606104), National Science Foundation of China (Grant number: 12222505), and the Strategic Priority Research Program of the Chinese Academy of Sciences (Grant number: XDA10010800), Y. M. and J. H. thank the sponsorship from the Yangyang Development Fund.\\

\bibliographystyle{elsarticle-num} 
\bibliography{bibfileTemplate}

\end{document}